\documentclass{statsoc}
\usepackage[utf8]{inputenc}

\usepackage{float}
\usepackage{amsmath}
\usepackage{amsfonts}
\usepackage{amssymb}
 
\usepackage{amsthm}
\usepackage{graphicx}
\usepackage{hyperref}
\usepackage{icomma}
\usepackage{latexsym}
\usepackage{textcomp}
\usepackage[all]{xy}
\usepackage{cancel}
\usepackage{color}
\usepackage{dsfont}
\usepackage{subfig}
\usepackage{caption}
\usepackage{dsfont}
\usepackage{extarrows}
\usepackage{shorttoc}
\usepackage{listings}
\usepackage[utf8]{inputenc}
\usepackage{lmodern}
\usepackage{xcolor}
\usepackage{moreverb}
\usepackage{longtable}

\usepackage[section]{placeins}

\usepackage{breakcites}

\usepackage{booktabs}
\usepackage{graphicx}

\RequirePackage[OT1]{fontenc}
\RequirePackage{amsthm,amsmath,natbib}
\RequirePackage{hypernat}

\usepackage{graphicx,color}

\usepackage{amssymb}

\usepackage{multirow}

 \newtheorem{assumption}{Assumption}[section]
     
     \newtheorem{theorem}{Theorem}[section]

\usepackage[T1]{fontenc} 
\usepackage[utf8]{inputenc}

\usepackage[ruled,vlined]{algorithm2e}
\SetKwProg{Init}{Init:}{}{}

\let\subparagraph\relax 

\usepackage{titlesec}
\titlespacing*{\section}{0pt}{12pt plus 0pt minus 2pt}{0pt plus 0pt minus 2pt}
\titlespacing*{\subsection}{0pt}{12pt plus 0pt minus 2pt}{0pt plus 0pt minus 2pt}
\titlespacing*{\align}{0pt}{12pt plus 4pt minus 2pt}{0pt plus 2pt minus 2pt}

\usepackage[a4paper]{geometry}

\let\OLDthebibliography\thebibliography
\renewcommand\thebibliography[1]{
  \OLDthebibliography{#1}
  \setlength{\parskip}{0pt}
  \setlength{\itemsep}{0pt plus 0.3ex}
}

\title[Testing for preferential sampling]{A fast Monte Carlo test for preferential sampling}
\author[J. Watson]{Joe Watson}
\address{University of British Columbia,
Vancouver,
Canada.}
\email{joe.watson@stat.ubc.ca}

\SetArgSty{textnormal}

\begin{document}

\maketitle

\begin{abstract}
The preferential sampling of locations chosen to observe a spatio-temporal process has been identified as a major problem across multiple fields. Predictions of the process can be severely biased when standard statistical methodologies are applied to preferentially sampled data without adjustment. Currently, methods that can adjust for preferential sampling are rarely implemented in the software packages most popular with researchers. Furthermore, they are technically demanding to design and fit. This paper presents a fast and intuitive Monte Carlo test for detecting preferential sampling. The test can be applied across a wide range of data types. Importantly, the method can also help with the discovery of a set of informative covariates that can sufficiently control for the preferential sampling. The discovery of these covariates can justify continued use of standard methodologies. A thorough simulation study is presented to demonstrate both the power and validity of the test in various data settings. The test is shown to attain high power for non-Gaussian data with sample sizes as low as 50.  Finally, two previously-published case studies are revisited and new insights into the nature of the informative sampling are gained. The test can be implemented with the R package \textit{PStestR}.
\end{abstract}

\keywords{Preferential sampling; Spatio-temporal statistics; Point processes; Spatial Statistics; Environmental monitoring; Geostatistics}

\section{Introduction}

This paper concerns preferential sampling (PS), where the locations selected to monitor a spatio--temporal process  
$\mu_{st},~s\in {\cal S},~t\in {\cal {T}}$, depend stochastically on the process they are measuring. PS is a special case of response--biased sampling. The space--time point is defined as $(\textbf{s}, t)\in {\cal S} \times{ {\cal{T}}}$. $\cal S$ denotes the spatial domain of interest and $\cal T$ denotes the temporal domain. Purely spatial processes (i.e. when ${\cal{T}}$ is a singleton) is a special case.

 To gain understanding of the process $\mu_{s, t}$, a set of time points $T \subset {\cal T}$ at which to observe $\mu_{s, t}$ are selected. Then, for each $t \in T$, a set of $n_t$ sampling locations $S_t \subset {\cal S}$ are chosen. Generally, the temporal domain ${\cal T}$ is a finite set, with $\mu$ a time--averaged quantity for practical reasons. Typically, $\mu_{s, t}$ is not observed directly and instead a noisy observation $Y_{s, t}$ is taken instead. The noise could be due to the presence of measurement error (i.e. the nugget effect) or other factors. $S_t$ may represent a set of points in space (i.e. $S_t = (\textbf{s}_i \in {\cal{ S}})_{i=1}^{n_t}$), or a set of well-defined areal units (i.e. $S_t = (\textbf{A}_i \subset {\cal{ S}})_{i=1}^{n_t}$). In this paper, these two cases are referred to as the geostatistical and discrete spatial settings respectively. In the latter case, observations will generally represent spatial--averages of $\mu_{s, t}$. 
 
 Difficulties arise with the estimation of $\mu_{s, t}$ when $S_t$ are selected in a preferential way. This is because most statistical methods for modeling spatio-temporal data, condition on the locations as fixed \citep{diggle2010geostatistical, cressie2015statistics}. Such models assume that locations were selected under
complete spatial randomness. Departures from this assumption in the true sampling scheme can lead to large biases in the prediction of the process $\mu_{s, t}$ \citep{watson2018general}. Hereafter, models that consider the locations as fixed are referred to as `naive'.

Additionally, a set of covariates $\textbf{X}_{s, t}$ may exist that influence the choice of sampling locations $S_t$. These covariates may also be associated with the underlying process being modeled $\mu_{s, t}$. When this occurs, including the neccessary $\textbf{X}_{s, t}$ in a `naive' regression model for $\mu_{s, t}$ may partially remove the deleterious effects of PS on the spatio-temporal prediction of $\mu_{s, t}$ \citep{gelfand2012effect}. This becomes a regression-adjustment approach to help correct for the departure from a complete spatial randomness sampling design for $S_t$. Covariates common to both the sampling process and $\mu_{s, t}$, are hereafter referred to as `informative' as in \citet{gelfand2012effect}. 
 
 
 Preferential sampling has been identified as a major concern across multiple fields. In ecology, PS may occur due to sightings data being comprised of opportunistic sightings or poorly-designed surveys. Observers frequently focus their efforts in areas where they expect to find the species, leading to PS \citep{fithian2015bias,watson2019general}. A consequence of this is that `naive' estimates of the geographical distribution of a species may be severely biased. Species' abundance estimates have also been shown to be affected \citep{pennino2019accounting}. PS should also be considered in the analysis of environmental data recorded from tagged animals. \citet{dinsdale2019modelling} demonstrated this with a case study using sea surface temperature recordings from tags attached to Elephant Seals in the Southern Indian ocean. The seals' preference for cooler waters led to biased `naive' spatial estimates of sea surface temperature.   
 
 In environmental statistics, the deleterious impacts of PS have been highlighted. For example, pollution concentration levels throughout $\cal S$ and $\cal T$ are commonly estimated using noisy observations, $Y_{s, t}$, recorded from environmental monitoring networks \citep{shaddick2018data}. Here, the locations of the monitors in a network $S_t$, may have been chosen in a preferential way
to meet specified objectives 
 \citep{schumacher1993using}. For example, urban air pollution monitoring sites are sometimes used for detecting noncompliance with air quality
standards \citep{ozone05,loperfido2008network}. In such settings, observations $Y_{s, t}$ will likely lead to overestimates of the overall levels of the air pollutant, $\mu_{s, t}$, throughout $\cal S$ and $\cal T$. These biased `naive' estimates $\hat{\mu}_{s, t}$ may then be unsuitable for assessing the impacts of $\mu_{s, t}$ on human health and welfare \citep{lee2015impact}.    

Previous PS tests have been developed for continuous spatial data, but limitations hinder their general use. Firstly,  \citet{schlather2004detecting} developed two Monte Carlo tests. Their null hypothesis assumes that the data are a realization of a \textit{random-field model}. They assume: the sampled point locations $S_t$ are a realization of a point process $\cal P$ on $\cal S$, the recorded values (called marks) of the points are the values of a realisation of a random field $\mu_{s, t}$ on $\cal S$, $\cal P$ and $\mu_{s, t}$ are independent processes. Independence here implies a non-preferential sampling mechanism. To detect departures from the null hypothesis, the authors define two characteristics of marked point process, denoted $E(d)$ and $V(d)$. These represent respectively the conditional expectation
and conditional variance of a mark, given that there exists another
point of the process at a distance $d$. These are chosen since under the null hypothesis E and V
should be constant. Monte Carlo tests are used to assess departures of
estimates of E and V from a constant function. This approach requires the assumption of Gaussian observations and hence does not generalise to non-continuous marks.

Next, \citet{guan2007test} developed an alternative simulation-free test for PS. Instead of fitting a parametric model for the marks, their approach instead divides the region $\cal S$ into non over-lapping subregions. These are assumed to be approximately independent, generating approximately IID replicates of the test statistic. The spatial range of $\mu_{s, t}$ can be thought of as representing the inter-point distance required for two observations of $\mu_{s, t}$ to be approximately independent. Finding a suitable set of subregions required for their test may prove a challenge when the spatial range of the correlation of $\mu_{s, t}$ is large relative to the size of $\cal S$.  Furthermore, this test requires very large sample sizes; their application used a sample size of over 4000. 

For modeling PS directly, it is common to take a model-based approach. Approaches often simultaneously fit a model for the observation process, $Y_{s, t}$, with a model for the sampling process, ${\cal{P}}$, within a joint-model framework \citep{diggle2010geostatistical}. Linear combinations of any spatio-temporal latent effects used to describe $\mu_{s, t}$ are shared across the linear predictors of the two processes. This sharing of latent effects helps to capture any stochastic dependence that may exist between the two processes. A nonzero effect estimate of any of these linear combinations provides evidence that PS is present \citep{watson2018general}. Whilst this approach has been successfully applied to mitigate PS, the use of this approach to test for PS may be out of reach of many researchers. These joint models are currently not implemented in many popular software packages, are computationally intensive to fit and can be difficult to design and interpret. Note that design-based approaches have also been introduced for specific scenarios \citep{zidek2014unbiasing}. Hereafter, the collection of spatio-temporal latent effects are denoted $\textbf{Z}_{s, t}$.


Due to the computational challenges of fitting joint models and the lack of generality of the current PS tests, PS appears to be often overlooked. Researchers may have non-Gaussian data, or have too small a sample size to perform either test. Consequently, without the ability to test for PS, researchers may then fit `naive' models to preferentially sampled data. The potential consequences of PS on their inferences may then be ignored. Fortunately, in many situations, a sufficient set of informative covariates $\textbf{X}_{s, t}$ may be available. These can help to control for the PS, without the need to fit joint models. Verifying the the existence of $\textbf{X}_{s, t}$ would allow researchers to confidently continue to use their preferred methodologies and packages.

This paper presents a computationally fast method for detecting PS. The algorithm for implementing the test is both intuitive and easy to program. The method primarily requires that the researcher be able to predict the values of $\mu_{s, t}$, and any latent spatio-temporal effect $Z_{s, t}$, throughout $\cal S$ and $\cal T$. Any preferred `naive' method can be used. The method is general in that it can test for PS in both the geostatistical and discrete spatial settings,  and can be used when the responses (marks) are non-Gaussian and even non-continuous. A general algorithm is provided for all settings. The test can also be adjusted for covariates. This allows researchers to discover a set of relevant covariates $\textbf{X}_{s, t}$ that can control the PS. 

Qualitatively, PS has a clear appearance in continuous spatial data. PS often appears as a clustering of locations chosen to observe $\mu_{s, t}$ in regions where one or more $Z_{s, t}$ is either high or low. The test in this paper directly targets this excess clustering. First, a suitable point process is fit to the observed locations to capture the true sampling process under the null hypothesis of no PS. Then, Monte Carlo (MC) realisations of the point process under the null are generated. The magnitude of correlation between the degree of clustering and the estimated values of $Z_{s, t}$ is computed for both the observed data and the MC realisations. If a stronger correlation is observed in the observed data compared with the MC samples, then evidence for PS has been found. The mean of the K nearest neighbours is our default recommendation to capture the degree of clustering as this quantity may also be used in the discrete spatial setting. In the discrete spatial setting, a Bernoulli sampling process is instead fit to a population of well-defined areal units under the null. A clustering of areal units chosen to observe $\mu_{s, t}$ in regions where $Z_{s, t}$ is either high or low indicates PS.  

The paper is organised as follows. Section 2 introduces the assumed marked point process data generating mechanism for the geostatistical setting. Then, the algorithm and properties of the PS test are described. Section 3 repeats the above for the discrete data setting. Section 4 demonstrates the power of the test to detect PS in a thorough simulation study. The joint effects of the: sample size, spatial smoothness of $\textbf{Z}_{s, t}$, spatio-temporal covariates $\textbf{X}_{s, t}$ and the magnitude of PS on the power of the test are discussed. Section 5 applies the test to two real datasets previously analysed in the literature. The PS test can be performed using the R package \textit{PStestR}.


\section{Preferential sampling in geostatistical data}


In continuous spatio-temporal settings, observations $Y_{s, t}$ are taken at a set of point locations $S_t$ within the study region $\cal S$ at each time step $t \in \cal T$. Standard approaches for modeling $\mu_{s, t}$ from a set of observations $Y_{s, t}$ include variogram analysis and kriging-based methods \citep{digglemodel}. These methods fall under the umbrella term of ``geostatistical methods'' and require the assumption that the locations chosen to observe the process $\mu_{s, t}$ were not preferentially sampled \citep{diggle2010geostatistical}.   

For the modeling of point patterns in space and time, spatio-temporal point processes are the standard statistical toolbox \citep{illian2008statistical, baddeley2015spatial}. This class of models will be used throughout our paper to explain the observed point patterns $S_t$ through time. Standard `naive' geostatistical methods require the assumption that the sampling process $\cal P$ generating the sampled locations $S_t$ be independent of the underlying spatio-temporal field $\mu_{s, t}$. This assumption implies no PS and simplifies the analysis greatly. Here, the point pattern $S_t$ and the marks $Y_{s, t}$ may be investigated separately using standard techniques. However, when this assumption is violated, the two processes must be considered together. Marked spatio-temporal point processes should be considered as a formal framework for such a data analysis \citep{schlather2004detecting}.

The PS test we are about to describe requires the following three assumptions. The final assumption describes the assumed characteristic behaviour of the PS.

\begin{assumption}\label{Assumption_1}
The PS is driven by some or all of the spatio-temporal latent effects $\textbf{Z}_{s, t}$.
\end{assumption}
\begin{assumption}\label{Assumption_2}
All latent effects $\textbf{Z}_{s, t}$ driving the PS are spatially `smooth enough' relative to both the size of the study region $|\cal S|$ and the number of locations chosen to sample the process $|S_t|$.
\end{assumption}
\begin{assumption}\label{Assumption_3}
The density of points within $S_t$ at space-time point $(\textbf{s}, t) \in \cal (S \times T)$ depends \textbf{monotonically} on the values of the components of $\textbf{Z}_{s, t}$ driving the PS.
\end{assumption}


Assumptions 2.1 - 2.3 imply that preferentially sampled data will appear as point patterns $S_t$ that are clustered in space for each $t \in T$. These clusters will focus around regions where relevant elements of $\textbf{Z}_{s, t}$ are especially high or low, depending on the direction of PS. The first inferential goal becomes the detection of monotonic associations between the degree of clustering throughout $\cal S \times T$ with the values of relevant $\textbf{Z}_{s, t}$. If PS is detected, then the second inferential objective becomes the determination of whether or not clustering can be explained by a set of informative covariates $\textbf{X}_{s, t}$. That objective is achieved if such a set removes all of the PS-associations. 

The ranked nearest neighbour distances between the sampling locations $S_t$ is proposed as a default choice to measure the magnitude of clustering. Following the recommendations of \citet{gignoux1999comparing}, edge-corrections for these distances are not considered within the Monte Carlo algorithm. Many other quantities can be chosen to capture local clustering and may be more suited for specific $S_t$ generating mechanisms. The PS test developed in this paper can easily be modified to use another quantity. The ranked nearest neighbour quantity is chosen for its generalisability across both discrete and continuous spatial settings.  

\subsection{Assumed model for preferential sampling}

Many spatio-temporal point processes have been developed, with each possessing fundamentally different properties. An appropriate choice for a given analysis depends upon the sampling protocols that generated $S_t$. For example, Gibbs point processes allow for second-order effects such as inter-point attraction and repulsion to exist between points. A limiting case is seen in the Hard Core process. This process does not allow for points to exist within a distance $R$, called the `range of interaction'. Cluster processes provide a class of point processes that describe the locations of `parent' points with a separate process from their `daughter' points \citep{baddeley2015spatial}. Many more processes exist and may prove useful in applications.   

The simplest class of spatio-temporally varying point processes is the inhomogeneous Poisson process (IPP hereafter) \citep{illian2008statistical}. The IPP is completely defined by its intensity function $\lambda(\textbf{s}, t)$. This is defined as the expected number of points per unit area and time immediately around $(\textbf{s}, t) \in \cal S \times T$. Let ${\cal{ S}} \subset{\mathds{R}}^2$. Define two disjoint space-time cubes $(A_1, T_1), (A_2, T_2) \subset{(\cal S, T)}$. Then the numbers of points that fall within the two space-time cubes $N(A_i, T_i)$ are independently Poisson distributed random variables with means: 

\begin{align}
    \Lambda(A_i, T_i) = \int_{A_i} \int_{T_i} \lambda(\textbf{s}, t) \textrm{d}t \textrm{d}\textbf{s}.
\end{align}

Gaussian processes $\textbf{Z}_{s, t}$ can be added to any linear predictor used to model the natural logarithm of $\lambda(\textbf{s}, t)$. $\lambda(\textbf{s}, t)$ then becomes a log-Gaussian random field and the process becomes known as a log-Gaussian Cox process (LGCP hereafter) \citep{simpson2016going}. LGCP models are especially useful for modelling point patterns when residual spatio-temporal correlations are expected to remain in the intensity, even after including any available covariates. In these cases, the $\textbf{Z}_{s, t}$ are given spatio-temporal correlation structures. The $\textbf{Z}_{s, t}$ are then referred to as spatio-temporal Gaussian random fields.  


A base LGCP model is now introduced for describing the sampling process of $S_t$ in many geostatistical settings. This model is very general. To ease the notational burden, only one latent effect is considered and denoted $Z_{s, t}$. This constraint can be relaxed. Note that when the sampling protocols generating $S_t$ clearly deviate from the LGCP assumption seen in (1), other point processes should be considered. Details of other processes are found in \citet{baddeley2015spatial} and \citet{ illian2008statistical}. 

With a slight change of notation, removing the subscripts to improve readability, let $Y(\textbf{s}, t)$ denote the observation process at location $\textbf{s} \in \cal S$ and time $t \in \cal T$. This may be of any type (e.g continuous, count, binary etc.). Let $\mu(\textbf{s}, t)$ denote the target spatio-temporal process and let $Z(\textbf{s}, t)$ denote the spatio-temporal latent Gaussian random field. As before, let $S_t$ denote the collection of sampled points at time $t \in \cal T$. The following data generating mechanism is now assumed:


\begin{align}
    [Y(\textbf{s}, t)|\textbf{s} \in S_t, Z(\textbf{s}, t)] &\sim f(\mu(\textbf{s}, t), \boldsymbol{\theta)} \\
    [S_t|Z(\textbf{s}, t)] &\sim \textrm{IPP}(\lambda(\textbf{s}, t)) \\
    g(\mu(\textbf{s}, t)) &= \beta_0 + \boldsymbol{\beta}^T\textbf{x}(\textbf{s}, t) + Z(\textbf{s}, t) \\
    \textrm{log}(\lambda(\textbf{s}, t)) &= \boldsymbol{\alpha}^T \textbf{w}(\textbf{s}, t) + \boldsymbol{\delta}(\textbf{x}(\textbf{s}, t)) + h(Z(\textbf{s}, t)) \\
    [Z(\textbf{s}, t)] &\sim \textrm{GP}(\textbf{0}, \boldsymbol{\Sigma}).
\end{align}

Square brackets denote random variables. In equation (2), $f$ represents the conditional probability distribution $Y(\textbf{s}, t)$, given the target latent spatio-temporal effect $Z(\textbf{s}, t)$, and given the location was sampled at time $t$ (i.e. $\textbf{s} \in S_t$). Values of $Y(\textbf{s}, t)$ are missing at all non-sampled locations. The link function $g$ describes the relationship between the linear predictor and the target spatio-temporal process $\mu(\textbf{s}, t)$. Thus, the model contains the popular class of generalised linear geostatistical models \citep{digglemodel}. The regression equation for $\mu(\textbf{s}, t)$ is specified in (4), with fixed covariates $\textbf{x}(\textbf{s}, t)$. 

In equation (3), the sampling process $\cal{P}$ is modeled as a LGCP with intensity function (5). Unique fixed covariates $\textbf{w}(\textbf{s}, t)$ and shared covariates $\textbf{x}(\textbf{s}, t)$ both describe the intensity. The shared covariates are transformed by functions $\boldsymbol{\delta}$ that may be nonlinear. Note that, conditional on $Z(\textbf{s}, t)$, $S_t$ is assumed to be a realisation from an IPP. The function $h$ allows for nonlinear transformations of the spatio-temporal Gaussian process $Z(\textbf{s}, t)$ to be included in (5). This specifies the nature of PS. 

The PS test developed in our paper is highly general. When $h$ is strictly monotonic, the primary goal of the test is to detect the monotonicity of $h$. The precise form of $h$ does not require specification. Suppose $h \equiv 0$ and at least one element of $\boldsymbol{\delta}$ is non-zero. The subset of covariates $\textbf{x}(\textbf{s}, t)$ corresponding to these nonzero elements provide a sufficient set of informative covariates required to control for PS. The second goal of the test is to correctly identify this subset of covariates. Note that the covariance matrix of the vector of $Z(\textbf{s}, t)$ values evaluated at $S_t$ is denoted $\Sigma$. This also requires estimation. Finally, $\boldsymbol{\theta}$ are hyperparameters to be estimated in the model. Both $\Sigma$ and $\boldsymbol{\theta}$ may be estimated using a maximum likelihood approach, or, given prior distributions, and then estimated under a Bayesian approach.

Including a sufficient set of informative covariates in a model for $\mu(\textbf{s}, t)$ should help to improve the prediction of $\mu(\textbf{s}, t)$ across $\cal S$ and $\cal T$, by reducing the deleterious impacts of PS on spatial prediction. However, it must be stressed that this approach is not a silver bullet; if the data ($S_t$) were instead collected via a complete spatial randomness sampling design, then the predictive accuracy of the fitted model would likely be improved \citep{gelfand2012effect}. Thus, these methods should be viewed only as a partial remedy for badly sampled data rather than as a justification for ignoring the need for good spatial design of networks and surveys.  

Finally, the conditional likelihood of the LGCP given $Z(\textbf{s}, t)$ is:

\begin{align}
    \pi \left(S_t | Z(\textbf{s}, t) \right) = \rm{exp}\left\{ |{\cal{ S}}| |T| - \int_{\cal{ S}} \int_T \lambda(\textbf{s}, t)dt d\textbf{s} \right\} \prod_{\textbf{s}_i \in S_t, t \in T}\lambda(\textbf{s}_i, t), \label{eq:LGCPlikelihood}
\end{align}
with $|{\cal{ S}}|$ being the area of the domain ${\cal{ S}}$ and $|T|$ being the length of the time set.
 
\subsection{Monte Carlo algorithm}

Assume the above data generating mechanism. A Monte Carlo algorithm is now designed for testing the null hypothesis that $h \equiv 0$, versus the alternative hypothesis that $h$ is a monotonic function of $Z$. Under the null hypothesis $h \equiv 0$, the observation and sampling processes are conditionally independent given $\textbf{x}(\textbf{s}, t)$. Thus, given $\textbf{x}(\textbf{s}, t)$, no associations are expected to exist between computable quantities from the fitted (null) IPP and estimates of $Z(\textbf{s}, t)$. 


Conversely, suppose that the null hypothesis is false. Specifically, let $h$ be a monotonic increasing function of $Z$. Point patterns $S_t$ from this data generating mechanism are expected to exhibit an excess of clustering in regions of high $Z(\textbf{s}, t)$, relative to that explained by the null model. This phenomenon is referred to as positive PS. Here, a positive association between the localised amount of clustering and estimated $Z(\textbf{s}, t)$ values would be expected. The converse holds when $h$ is a decreasing function. 

The primary challenge is defining what constitutes a `strong' association between estimates of $Z(\textbf{s}, t)$ and the computed quantities used to capture excess localised clustering. Positive spatio-temporal correlations are present in $\mu(\textbf{s}, t)$ due to $Z(\textbf{s}, t)$. This leads to non-standard sampling distributions for test statistics computed to capture association. Standard hypothesis tests of association (e.g.\ t-tests, rank-correlation tests etc.,) will have a type 1 error above the specified level due to the positive correlations. 

This is why Monte Carlo methods are used. An empirical p-value associated with any desired test statistic can be computed by sampling realisations from the assumed IPP under the null hypothesis (i.e. fixing $h \equiv 0$). The application generalises to any given dataset. Crucially, this procedure accounts for the nonstandard sampling distribution of the chosen test statistic in a natural way. The mean of the $K$ nearest neighbour distances from each observed point is our default choice of computable quantity. Small values of this quantity within a region, indicates the presence of clustering there. When $K=1$, this reduces to the nearest neighbour distance. For the default choice of test statistic, the Spearman's rank correlation coefficient between estimates of $Z(\textbf{s}, t)$ at locations $\textbf{s} \in S_t$ and the mean nearest neighbour distances is proposed. Unlike Pearson's correlation coefficient, Spearman's rank correlation measures the degree of monotonicity of $h$, instead of the degree of linearity. 

We now state the probability distribution function of the (spatial) distance from any point of $S_t$ to its nearest neighbouring point for the IPP data model. Let ($\textbf{s}, t) \in \cal S \times T$ define a reference space time point and let $T$ be the time interval $(t, t_T)$ of interest respectively. Next, define $b(\textbf{s}, r)$ as the ball of radius $r$ centered at $\textbf{s}$. Let $\lambda_{\textrm{IPP}}(\textbf{s}, t)$ once again denote the intensity function for the assumed IPP model under the null hypothesis $h \equiv 0$. The following assumption is made:

\begin{assumption}\label{assumption_IPP_2}
$h \equiv 0$
\end{assumption}

\begin{theorem}
Under Assumption 2.4 and the above data generating mechanism, the probability that the nearest point of $S_\tau$ from $(\textbf{s}, t)$, lies within a spatial distance $r$, at some time $\tau \in T$ is: 
\begin{align}
 1 - \textnormal{exp}\left(-\int_{b(\textbf{s}, r)} \int_{T} \lambda_{\textnormal{IPP}}(\boldsymbol{\omega}, \tau) \textnormal{d}\tau \textnormal{d}\boldsymbol{\omega} \right) =  1 - \textnormal{exp}\{-\Lambda(b(\textbf{s}, r), T)\}.
\end{align}
\end{theorem}

Equation (8) gives us the following intuitive result for IPPs. When $h \equiv 0$, the expected nearest neighbour distances are lower in regions of high intensity (i.e where $\lambda_{\textrm{IPP}}(\textbf{s}, t)$ is high). This is to be expected - the intensity function at $(\textbf{s}, t)$ precisely defines the expected density of points immediately around $(\textbf{s}, t)$.  

The result has also been derived when $h \neq 0$, under the alternative LGCP model. \citet{coeurjolly2017palm} derived the Palm distribution for LGCPs. The authors showed the remarkable result that conditional on a single point in $S_t$ lying at location $(\textbf{s}, t) \in \cal S \times \cal T$, the remaining points of $S_t$ are also a LGCP. The second order joint intensity function describing this process can be thought of as representing the intensity, conditional on $(\textbf{s}, t) \in S_t$. Using the authors' result, this remains a log-Gaussian random field, with the conditional process differing only in the mean.

\begin{assumption}\label{assumption_LGCP_1}
$h \neq 0$, with $h$ a positive monotonic function.
\end{assumption}
\begin{assumption}\label{assumption_LGCP_2}
The covariance is strictly non-negative, i.e. $\Sigma(\cdot , \cdot) \geq 0$
\end{assumption}
\begin{assumption}\label{assumption_LGCP_3}
$Z$ is stationary and isotropic: $\Sigma(\textbf{s} - \boldsymbol{\omega}, t - \tau) = \sigma^2_Z R(|| \textbf{s} - \boldsymbol{\omega} ||, || t - \tau ||)$
\end{assumption}
\begin{assumption}\label{assumption_LGCP_4}
The correlation function decays monotonically as the distance from the conditioning point $\textbf{s}$ increases: $\sigma^2_Z R(|| \textbf{s} - \boldsymbol{\omega} ||, || t - \tau ||) \geq \sigma^2_Z R(|| \textbf{s} - \boldsymbol{\omega} || + \delta, || t - \tau ||) \hspace{0.1cm} \forall \delta > 0$ .
\end{assumption}
\begin{assumption}\label{assumption_LGCP_5}
The correlation function decays monotonically as the distance from the conditioning time $t$ increases: $\sigma^2_Z R(|| \textbf{s} - \boldsymbol{\omega} ||, || t - \tau ||) \geq \sigma^2_Z R(|| \textbf{s} - \boldsymbol{\omega} ||, || t - \tau || + \delta) \hspace{0.1cm} \forall \delta > 0$ .
\end{assumption}

\begin{theorem}
Under Assumption 2.5 and the above data generating mechanism, the second order joint intensity function $\lambda^{(2)}_{\textbf{s}}(\cdot, \cdot)$ at $(\boldsymbol{\omega}, \tau$) is:

\begin{align}
\textnormal{log}(\lambda^{(2)}_{\textbf{s}}(\boldsymbol{\omega}, \tau)) &= \boldsymbol{\alpha}^T \textbf{w}(\boldsymbol{\omega}, \tau) + \boldsymbol{\delta}(\textbf{x}(\boldsymbol{\omega}, \tau)) + h(Z(\boldsymbol{\omega}, \tau)) + \Sigma(\textbf{s} - \boldsymbol{\omega}, t - \tau). 
\end{align}

Then, using the law of total expectation, the probability that the nearest point from $(\textbf{s}, t)$, lies within distance $r$, at some time $\tau \in T$ is:

\begin{align}
    1 - E_Z \left[ \textnormal{exp} \left( - \int_{b(\textbf{s}, r)} \int_T \lambda_{\textnormal{IPP}}(\boldsymbol{\omega}, \tau) \left\{ \textnormal{exp}\left[h(Z(\boldsymbol{\omega}, \tau)) + \Sigma(\textbf{s} - \boldsymbol{\omega}, t - \tau)\right] \right\} \textnormal{d}\tau \textnormal{d}\boldsymbol{\omega} \right) \right].
\end{align}

\end{theorem}

Thus, within the linear predictor, the only change from the original LGCP intensity (5) is the addition of the term $\Sigma(\textbf{s} - \boldsymbol{\omega}, t - \tau)$. This is the covariance function between the conditioning point $(\textbf{s}, t)$ and the space-time point $(\boldsymbol{\omega}, \tau)$. Under Assumptions 2.5 - 2.9, the intensity immediately around $(\textbf{s}, t)$ will always be higher. Contrast this with the IPP. Here, the knowledge of a point of $S_t$ existing at $(\textbf{s}, t)$ does not affect the intensity immediately around $(\textbf{s}, t)$. Note that the conditional intensity of the point process monotonically increases with $h(Z)$ and $\Sigma(\cdot, \cdot)$. 

Assumptions 2.6 - 2.9 are commonly made in practice. They imply that the latent process will be expected to be more similar at two space-time locations that are `close together' than two space-time locations that are `far apart'. Popular choices of correlation functions include the Matern correlation function across space \citep{digglemodel} and autoregressive correlation functions across time. Spatio-temporal correlation functions are often defined to be the products of these spatial and temporal functions for computational simplicity \citep{blangiardo2015spatial}. Under these models, the correlation often becomes negligible at spatial (temporal) distances greater than some value, often called the spatial (temporal) range.  

Under Assumptions 2.5 - 2.9, equation (10) now helps to explain the suitability of our choice of nearest neighbour distance to capture the excess clustering. Firstly, suppose the conditioning point $(\textbf{s}, t)$ is in a region where the latent effect is above average (i.e.\ $Z(\textbf{s}, t) > 0$). Here, we see that the expected nearest neighbour distance from $(\textbf{s}, t)$ decreases monotonically as: i) $h(Z)$ increases and ii) the correlation function $R(\cdot, \cdot)$ increases. Condition ii) implies that the nearest neighbour distances will decrease as the size of the spatial and temporal ranges increase. Note that the converse results hold when $(\textbf{s}, t)$ is in a region of low $Z(\textbf{s, t})$. 

Thus we have shown the result we wanted. Under Assumptions 2.6 - 2.9, with $h$ monotonic in \textbf{either} direction, a monotonic association is expected between the nearest neighbour distances between the observed points $(\textbf{s}, t) \in S_t$ and the values of $Z$ at $S_t$. This monotonic association should be captured with our rank correlation test statistic when $K = 1$. Conversely, under Assumption 2.4, no association is expected, so long as the fitted null IPP is correctly specified. In this case, the observed test statistic should be no more extreme than the Monte Carlo realisations.


To define the test requires some additional notation. Let $T$ be a finite set of time intervals. Let $n_t$ denote the observed number of points $ \textbf{s}_{i, t} \in S_t \subset \cal S$ at time $t \in T$. Let $N_{i, t}(K)$ denote the set of K nearest indices from each point $\textbf{s}_{i, t} : i \in \{1,...,n_t\}$. Let $\hat{Z}(\textbf{s}, t)$ denote the estimate of $Z(\textbf{s}, t)$. Define the superscript above each of these quantities $m$ as the index of the Monte Carlo sample $m \in \{1,...,M\}$. Thus $N_{i, t}^{m}(K) : i \in \{1,...n^m_t\}, m \in \{1,...,M\}$ denotes the set of K nearest indices from point $\textbf{s}_{i, t}^{m}$ in the $m^{th}$ Monte Carlo sample $S_{t}^m$. Finally, let $\Bar{D}_{i, t}(K), \Bar{D}_{i, t}^{m}(K)$ denote the mean of the distances to the $K$ nearest points from point $i$ at time $t$ in the original data and the $m^{th}$ Monte Carlo sampled point pattern respectively. Thus:

\begin{align}
    \Bar{D}_{i, t}^{m}(K) = \frac{1}{K} \sum_{j \in N_{i, t}^{m}(K)} || \textbf{s}^{m}_{i, t} - \textbf{s}^{m}_{j, t} ||.
\end{align}

The terms $NN_{k, t}$ and $NN^m_{k, t}$ are defined to be the vectors of length $n_t$ and $n_t^m$ containing the values of $\Bar{D}_{i, t}(K)$ and $\Bar{D}_{i, t}^{m}(K)$ respectively. When calculating (9) in the original dataset, simply drop the $m$ superscripts. 

\vspace{1pt}

\begin{algorithm}[H]
\DontPrintSemicolon
 \caption{Monte Carlo $NN$ test for PS in geostatistical data}
\SetAlgoLined
\KwData{\\ Observations $\mathbf{y}(\mathbf{s}, t)$ for ($\mathbf{s}, t) \in (S_t \times T) \subset  (\cal S \times T)$ \\ Covariates $\{\mathbf{w}(\mathbf{s}, t)$, $\mathbf{x}(\mathbf{s}, t)\}$ for ($\mathbf{s}, t) \in ({\cal{S}} \times T)$ }
\KwResult{\\ Empirical p-value for the test $h \equiv 0$ vs. $h$ monotonic }
\Begin{
Fit a model for (2) using a preferred method\;
Produce estimates $\hat{Z}(\mathbf{s}, t)$ throughout $\cal S, T$\;
Compute the $NN_{k, t}$ values $\Bar{D}_{i, t}(K)$\;
Evaluate $\hat{Z}(\mathbf{s}, t)$ at locations $S_{t}$\; 
 Compute the rank correlations $\rho_t$ between $\hat{Z}(\mathbf{s}_i, t)$ and $\Bar{D}_{i, t}(K)$\;
Fit the chosen point process model with $h \equiv 0$ in (5)\; 
 Fix $m = 1$\;
\While{$m \leq M$}{
 Sample $n_t^m$ locations $S_{t}^m$ from the fitted model for $t \in T$\;
  Compute the $NN_{k, t}$ values $\Bar{D}_{i, t}^{m}(K)$\;
  Compute $\hat{Z}(\mathbf{s}, t)$ at locations $S_{t}^m$\;
  Compute the rank correlations $\rho_{t}^{m}$ between $\hat{Z}(\mathbf{s}^m_i, t)$ and $\Bar{D}_{i, t}^{m}(K)$\;
  \eIf{$m = M$}{
   return the empirical p-values of either pointwise or rank envelope tests using $\rho_t$ and $\rho_t^m$. \;
   }{
   $m \gets m+1$ \;
  }
 }
 }
\end{algorithm}

\vspace{1pt}

The Monte Carlo algorithm, referred to as the $NN$ test hereafter, is defined above in Algorithm 1. It can now be summarised as follows. First, fit the assumed models (2) and (4) for both $Y(\textbf{s}, t)$ and $S_t$. Next, estimate $\hat{Z}(\textbf{s}, t)$ throughout $\cal S \times T$ and compute the averaged K nearest neighbour distances $NN_{k,t}$. Using the estimates $\hat{Z}(\textbf{s}, t)$ at $S_t$ and $NN_{k, t}$, compute the Spearman's rank correlation coefficient $\rho_t$ between them for each $t \in T$. This is the observed test statistic.  

Next, sample $M$ realisations, $S_t^m : m \in \{1,...,M\}$, from the fitted point process model (4). For each of the $M$ realisations, repeat the procedure. Compute the distances $NN_{k, t}^m$ and estimate the values $\hat{Z}(\textbf{s}, t)$ at $S_t^m$ to obtain $\rho_t^m : m \in \{1,...,M\}$. Finally, compute the desired empirical p-value. For the pointwise tests, simply evaluate the proportion of the Monte Carlo-sampled $\rho_t^m$ that are more extreme than $\rho_t$. For Monte Carlo envelope tests that do not suffer from the problems of multiple testing, refer to \citet{mrkvivcka2017multiple, myllymaki2017global}.

\subsection{Discussion}
The values of the latent field $Z$ are not known and must be estimated. The power of the test to detect PS, may depend upon the suitability of the method used to produce estimates $\hat{Z}(\textbf{s}, t)$. Likelihood-based approaches for fitting the above observation model (i.e. components (1), (3) and (5)) have been shown to have many nice properties. Asymptotic consistency has been proven under the null hypothesis that $h \equiv 0$ for certain choices of $f$. These include Bernoulli \citep{ghosal2006posterior} and Gaussian \citep{choi2007posterior} likelihoods. Such results help to justify the suitability of the method. Furthermore, under the null hypothesis $h \equiv 0$, some `naive' approaches even produce unbiased estimates of $Z(\textbf{s}, t)$. For example, under the above data generating mechanism, with $f$ the normal distribution and $g$ the identity function, the Gaussian process regression is the best linear unbiased predictor for $Z$ \citep{cressie1992statistics}. 

Other choices of a computable quantity for capturing spatial clustering can be made. These may be more suitable for certain data generating mechanisms of $S_t$. However, few choices are as generalisable across both continuous and discrete spatial data. For continuous spatial data, smoothed estimates of the residual measure from the fitted (null) point process, evaluated at the points $\textbf{s} \in S_t$ may be suitable. However, this depends upon two tuning parameters: the details of the discretisation method chosen to approximate the likelihood and the choice of the bandwidth used to smooth the estimated values.  

The nearest neighbour method does not suffer these drawbacks and has many desirable properties. It generalises across both continuous and discrete spatial settings. Secondly, different choices of $K$ can lead to improved powers to detect PS under different sampling processes. For example, when the spatial scales of clusters within $S_t$ are very small, and hence when clustering is very localised to only a few points per cluster, smaller $K$ may lead to improvements in power. This is because larger values of $K$ may `smooth over' any clustering. Conversely, when clusters are large in spatial scale, with each cluster being comprised of several points, the power may be improved with larger choices of $K$. Here, the additional smoothing can reduce the variance of the computed test statistic. The final benefit of the nearest neighbour quantity is that the distances can be computed exactly, with values not dependent upon any choice of computational approximation.

In some applications it may be suitable to fix the sample size across the Monte Carlo samples (i.e. $n_t^m \equiv n_t$). For example, regulatory standards may dictate the required number of monitoring sites $n_t$. The assumption of conditional independence between the sampled locations under the null IPP makes this especially easy.


Strictly speaking, since in practice the values of $Z(\textbf{s}, t)$ and the parameters in (5) are not known and are only estimates, the plug-in test of Algorithm 1 will be invalid. This is because the null hypothesis is composite \citep{baddeley2017two}. However, tests which ignore the effects of parameter estimation will tend to be conservative in most cases. A loss of power is typically the price to pay \citep{dao2014monte}. Whilst the method introduced by \citet{dao2014monte} can be used to ensure the test attains nominal type 1 error, the required nested Monte Carlo simulations dramatically slows down the implementation. In the simulation studies of Section 4, the test defined in Algorithm 1 is found to be conservative across all tested simulation settings. Thus we do not consider this matter further.

P-values have come under increasing criticism recently (see \citet{wasserstein2016asa} and references within). Indeed, the computation of an empirical p-value alone to identify the binary presence/absence of PS within a dataset has its flaws. For example, it does not help to quantify the potential magnitude of the biasing effects that the PS may have on the spatial prediction of $\mu(\textbf{s}, t)$. Furthermore, as with all p-values, it is easy to fall victim to the p-value fallacy. A given p-value does not provide much information on its own. A value close to 0.05 neither provides strong evidence in favour of the alternative hypothesis vs. the null hypothesis, nor implies that the frequentist error probability is close to 0.05 \citep{sellke2001calibration}. Additional steps must be taken to make such inferences, such as the use of the calibrations introduced by  \citet{sellke2001calibration}. In summary, any reported p-values from the tests outlined in algorithms 1 and 2 must be used with care.  

Assuming the correct data generating mechanism is specified and the true parameters are known, the test will be exact regardless of how small $M$ is chosen. For testing at the 5\% significance level, $M$ could be chosen as low as 19. However, this comes at a cost of power, with the loss of power proportional to $1/M$ \citep{davidson2000bootstrap}. Furthermore, a small $M$ implies a high standard error of the empirical p-value. This leads to a test whose outcome is heavily dependent on the precise sequence of random numbers used to implement the algorithm. To alleviate these concerns, $M$ should be chosen as large as is computationally feasible.


\section{Preferential sampling in the discrete spatial data setting}

In the discrete spatial data setting, observations are taken across a set of areal units $S_t$ within the study region $\cal S$. Examples of areal units include electoral districts and large survey transects. The sizes of these areal units may be irregular, and are assumed known. It is also assumed that the full population of all possible areal units that were available for sampling at each $t \in T$ is known. This population is denoted $P_t$. A binary process is fit to emulate the true sampling process. The choice between a Bernoulli and Binomial model depends on whether or not the constraint $n_t^m = n_t$ is implemented. Once again, a Monte Carlo approach is taken for testing for PS.

\subsection{Assumed model for preferential sampling}

Given the population of $n_t$ areal units available at time $t$, denoted $P_t = \{ \textbf{A}_{i, t} \subset {\cal {S}} : i \in \{1,...,n_t\} \}$, define the site-selection indicator variables as follows. Let $R_{i}(t)$ denote the indicator random variable that the $i^{th}$ areal unit in $P_t$ is selected at time $t$. Then the collection of sampled areal units $S_t \subset P_t$ at each time $t \in T$, is simply the subset of the population of areal unit whose indicator variables take value 1 (i.e. $S_t = \{\textbf{A}_{i, t} \subset P_t : R_i(t) = 1 \}$). 

Next, define $\Bar{\textbf{w}}(\textbf{A}, t)$ to be the fixed spatio-temporal covariates for the indicator selection process at areal unit $\textbf{A}$. These will typically be areal-aggregate or areal-count values. Similarly, $\Bar{Z}(\textbf{A}, t)$ and $\Bar{\textbf{x}}(\textbf{A}_i, t)$ will typically be areal-aggregates of the underlying spatio-temporal process $Z(\textbf{s}, t)$ and the spatio-temporal covariates $\textbf{x}(\textbf{s}, t)$ respectively. In applications, $\Bar{Z}(\textbf{A}, t)$ will typically be modeled as a discrete spatio-temporal process on the areal unit scale instead of as a continuous process. Examples include the conditional autoregressive process and its spatio-temporal extensions \citep{besag1974spatial, blangiardo2015spatial}.

The same model form is assumed for the observation process $Y(\textbf{A}, t)$ in (1). The only change made is the spatial scale. Thus, the new sampling process $\cal P$ is defined as:

\begin{align}
    [R_i(t)|\Bar{Z}(\textbf{A}_i, t)] &\sim \textrm{Bernoulli}(p(\textbf{A}_i, t)) \\
    \textrm{logit}(p(\textbf{A}_i, t)) &= \boldsymbol{\alpha}^T \Bar{\textbf{w}}(\textbf{A}_i, t) + \boldsymbol{\delta}^T \Bar{\textbf{x}}(\textbf{A}_i, t) +  h(\Bar{Z}(\textbf{A}_i, t)).
\end{align}

For each time step $t \in T$, it is assumed that each of the areal units $\textbf{A}_i$ within the population $P_t$ has values $Y(\textbf{A}_i, t)$ sampled or not according to the outcomes of the independent Bernoulli trials defined in (12).

\subsection{Monte Carlo algorithm}

\begin{algorithm}[H]
\DontPrintSemicolon
 \caption{Monte Carlo PS test for discrete spatial data}
\SetAlgoLined
\KwData{\\ Observations $\mathbf{y}(\mathbf{A}_i, t)$ for ($\mathbf{A}_i, t) \subset (S_t \times T)$\\ 
Covariates $\Bar{\mathbf{w}}(\mathbf{A}_{i}, t)$, $\Bar{\mathbf{x}}(\mathbf{A}_i, t)$ for ($\mathbf{A}_i, t) \subset (P_t \times T)$  }
\KwResult{\\ Empirical p-value for the test $h \equiv 0$ vs. $h$ monotonic }
\Begin{
Fit a model for (1) using a preferred method\; 
Produce estimates $\hat{\Bar{Z}}(\mathbf{A}_i, t)$ across $P_t$ and $T$\;
Compute the $NN_{k, t}$ values $\Bar{D}_{i, t}(K)$\;
Compute $\hat{\Bar{Z}}(\mathbf{A}_i, t)$ at areal units $S_{t}$\;
Compute the rank correlations $\rho_{t}$\;
Fit the chosen Bernoulli model with $h \equiv 0$ in (9)\; 
Fix $m = 1$.
 \While{$m \leq M$}{
 Sample $n_t^m$ areal units $S_{t}^m$ from the fitted model for $t \in T$\;
  Compute the NN distance measure $\Bar{D}_{i, t}^{m}(K)$\;
  Compute $\hat{\Bar{Z}}(\mathbf{A}_i, t)$ at areal units $S_{t}^m$\;
  Compute the rank correlations $\rho_{t}^{m}$\;
  \eIf{$m = M$}{
   return the empirical p-values of either pointwise or rank envelope tests using $\rho_t$ and $\rho_t^m$.
   }{
   $m \gets m+1$\;
  }
 }
 }
\end{algorithm}

\vspace{1pt}

All of the same assumptions and issues outlined earlier carry over to the discrete spatial setting. Once again, $Z(\textbf{A}_i, t)$ must be spatially smooth across the areal units and estimates of $\Bar{Z}(\textbf{A}_i, t)$ must be available at each of the areal units $\textbf{A}_i$ in the population $S_t$ at each time $t \in T$. Nearest neighbour distances between areal units within $S_t$ can once again be used. Such distances can be defined relative to the areal unit-centroids or otherwise. Strictly speaking, the PS is no longer seen as a clustering of the point pattern around high (or low) values of $\Bar{Z}(\textbf{A}_i, t)$. Instead, the PS takes the form of a clustering of the areal units with complete data around high (or low) values of $\Bar{Z}(\textbf{A}_i, t)$. The procedure is defined in Algorithm 2 above. 

\section{Simulation Study}

This section summarises the key results of an investigation into the performance of the $NN$ test. The power of the test is demonstrated across a range of simulated data settings. A more thorough treatment of the simulation study is provided in the supplementary material. All computations involving point processes were performed using the \textit{spatstat} package \citep{baddeley2015spatial}. 

The following data generating mechanism is chosen for the Gaussian response simulation study:

\begin{align}
    [Y(\textbf{s})|Z(\textbf{s})] &= Z(\textbf{s}) \\
    [S|Z(\textbf{s})] &\sim \textrm{IPP}(\lambda(\textbf{s})) \\
    \textrm{log}(\lambda(\textbf{s})) &= \alpha_0 + \alpha_1 w(\textbf{s}) + \gamma Z(\textbf{s}) \\
    [Z(\textbf{s})] &\sim \textrm{GP}(\textbf{0}, \Sigma).
\end{align}

The simulated data are in the purely spatial setting (i.e. |T| = 1), with the $Y(\textbf{s})$ specified as noise-free observations of $Z(\textbf{s})$. $Z(\textbf{s})$ is a realisation of a mean-zero Gaussian process with Matern covariance matrix $\Sigma$. The Matern roughness parameter $\nu$ is set to 1 and the standard deviation of $Z(\textbf{s})$ is fixed at 1. The spatial range $\rho_Z$ of the process is adjusted. The spatial range is defined here to be the distance at which the spatial correlation drops below 0.1. A larger $\rho_Z$ implies the process has a greater spatial smoothness (i.e.\ a lower frequency).  

The sampled locations are generated from a LGCP with a single covariate $w(\textbf{s})$. The number of points is fixed equal to $n$, and thus the true process is a Binomial point process. The parameter $\gamma$ determines the magnitude of PS, with $\gamma = 0$ corresponding to the null IPP model of no PS. Again, $w(\textbf{s})$ is an independent realisation of another Gaussian process with Matern covariance function. Both the roughness $\nu$ and the standard deviation are again fixed at 1, but the range parameter $\rho_w$ is varied independently from $\rho_Z$. The values of $w(\textbf{s})$ are assumed known throughout $\cal S$. The parameter $\alpha_1$ determines the effect of the covariate on the intensity $\lambda(\textbf{s})$.

The $NN$ test is performed at the 5\% significance level using 19 Monte Carlo samples (i.e. $M = 19$). $M = 19$ implies that these results provide a lower bound on the power of the test \citep{davidson2000bootstrap}; a higher power would have been attained had a larger $M$ value been chosen. All tests are performed with the two-sided alternative hypothesis that $h$ is a monotonic function of $Z$. Each experimental setting is repeated 200 times. In the study, all combinations of the following parameters are evaluated:

\begin{itemize}
    \item Sample size $n \in \{50, 100, 250\}$
    \item PS magnitude $\gamma \in \{0, 1, 2\}$
    \item Covariate effect $\alpha_1 \in \{0, 1\}$
    \item Spatial range of $Z$, $\rho_Z \in \{1.00, 0.20, 0.02\}$
    \item Spatial range of $w$, $\rho_w \in \{1.00, 0.02\}$
    \item Number of nearest neighbour distances $K \in \{1,...,15\}$
\end{itemize}

Along with the $NN$ test outlined in Algorithm 1, a Monte Carlo test using estimates of the raw residuals of the assumed IPP under the null hypothesis is also computed. This time the rank correlation between the estimates of $Z(\textbf{s})$ and the estimated residuals is the test statistic. As before, both sets of estimates are only evaluated at the point locations. The estimated residual values are simply kernel smoothed raw-residuals. An edge-corrected Gaussian kernel is used with bandwidth selected using leave-one-out cross-validation. We refer to this test as the residual test hereafter. It is interesting to assess the relative performance of the $NN$ test, given its generality across all point processes and to the discrete spatial setting. We now summarise the results of the study.  
 
\begin{figure}[ht!]
    \centering
    \includegraphics[scale=0.6]{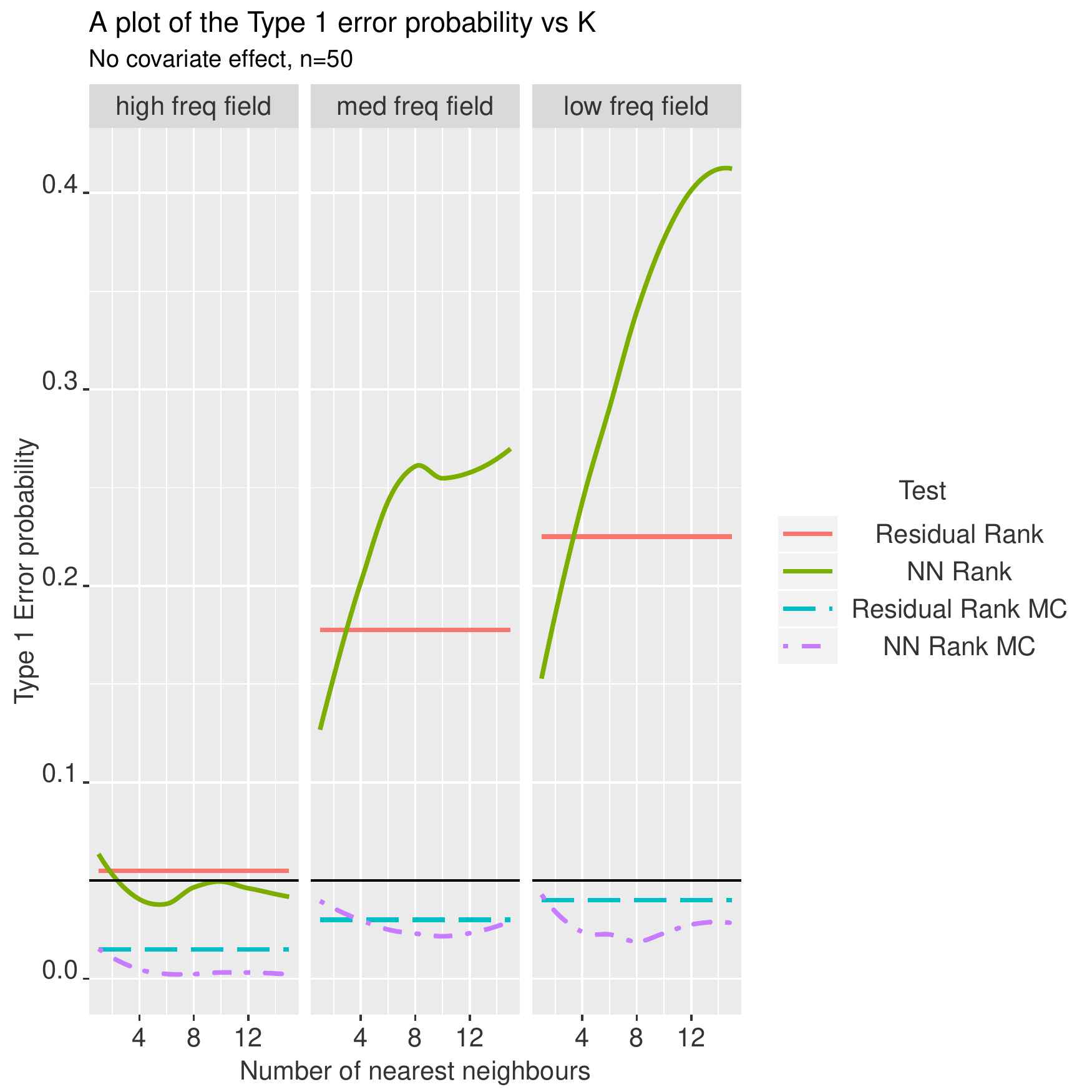}
    \caption{A plot of the Type 1 error for four tests. The three boxes show the results for $\rho_Z \in \{0.02, 0.2, 1\}$, from left to right respectively for a sample size of 50. The two `Residual' tests are computed using the kernel-density smoothed values of the residuals from the fitted homogeneous Poisson processes. Leave-one-out cross validation was used to select the bandwidth. The `NN' tests are those based on the $K$ nearest neighbour values. The suffix `MC' denotes the test has been computed from Monte Carlo realisations of the fitted point process. A black line is plotted at the type 1 error level of 0.05 to indicate the target value.}
    \label{fig:Type1_n50}
\end{figure}

First, our investigation strongly suggests that the type 1 error is bounded above by the nominal level. This is seen across all simulation settings when the null hypothesis is true. Thus, it appears that the computationally costly nested Monte Carlo approach of \citet{dao2014monte} need not be used, except in the interest of improving the power of the test. Fig.\ \ref{fig:Type1_n50} shows the results for $n=50$ in the simplest setting without covariate effects (i.e. $\alpha_1 = 0$) or PS effects (i.e. $\gamma = 0$). The spatial range $\rho_Z$ is changed and the test is performed across different value of $K$. It is apparent that both Monte Carlo tests attain Type 1 error at or below the 5\% level. For comparison, the two standard simulation-free rank correlation tests attain a Type 1 error well above the 5\% level. The error of these tests increases dramatically with $\rho_Z$ because they ignore the spatial correlation and hence the non-standard sampling distribution of the test statistic. We omit the simulation-free results hereafter.

\begin{figure}[ht!]
    \centering
    \includegraphics[scale=0.6]{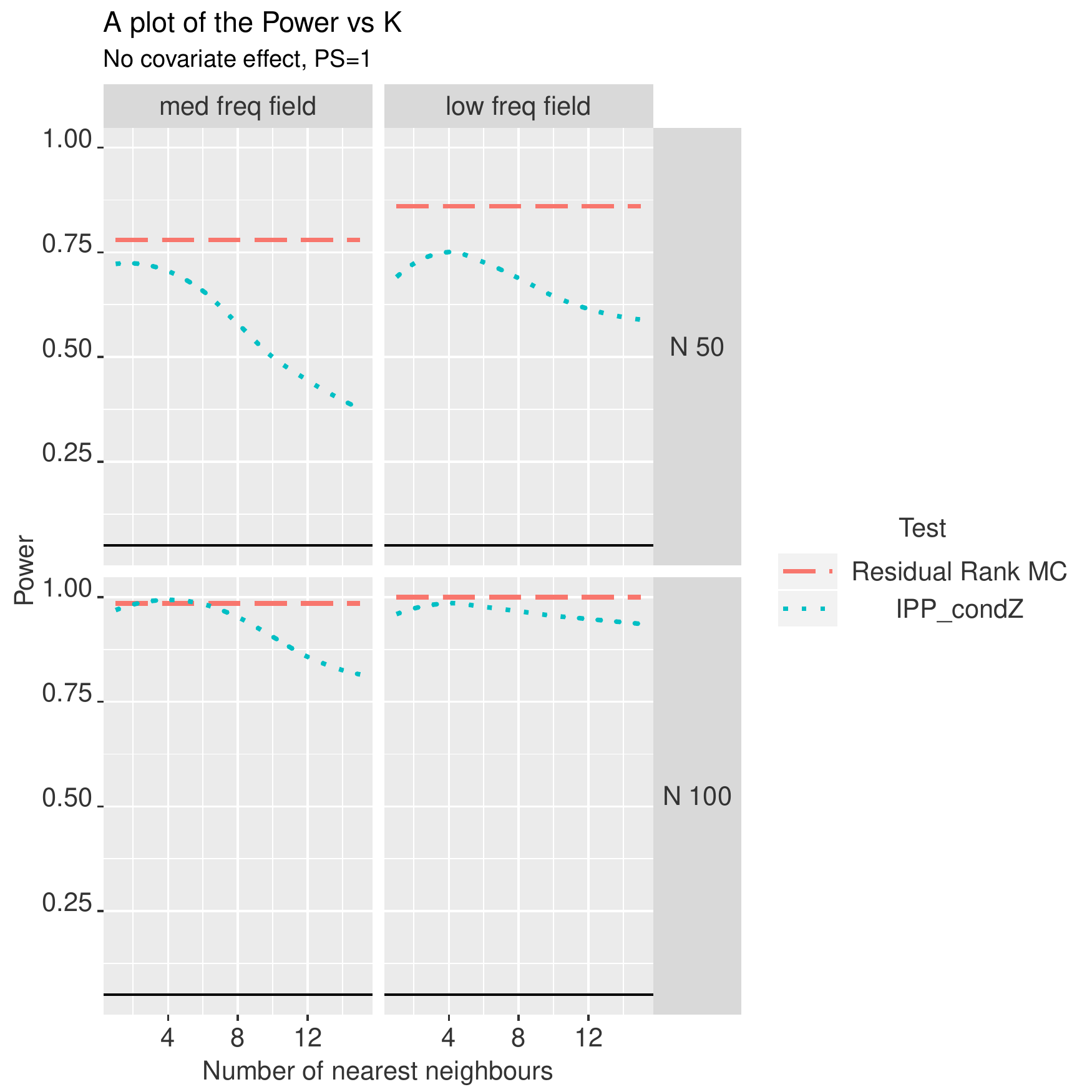}
    \caption{A plot of the Power for two tests when the PS parameter $\gamma$ equals 1. The two columns show the results for $\rho_Z \in {0.2, 1}$, from left to right respectively. The two rows show results for a sample size of 50 and 100 respectively. The `Residual' tests are computed using the kernel-density smoothed values of the residuals from the fitted homogeneous Poisson processes. Leave-one-out cross validation was used to select the bandwidth. The `NN' test is based on the $K$ nearest neighbour values. The suffix `MC' denotes the test has been computed from Monte Carlo realisations of the fitted point process.}
    \label{fig:Power_n50}
\end{figure}

Next, the power is assessed. Across all the simulation settings, the power improved with increasing spatial range $\rho_Z$. This is in agreement with the earlier results (9) and (10). $Z(\textbf{s}, t)$ must be spatially-smooth to achieve high power. Furthermore, the power of the $NN$ test is found to be sensitive to the choice of $K$ value. Optimal choice of $K$ depends upon both the spatial range of $Z$ and the sample size. Larger values of both implies that a larger value of $K$ should be chosen. Fig.\ \ref{fig:Power_n50} shows the results for the setting where no covariate effects exist (i.e. $\alpha_1 = 0$), but where moderate positive preferential sampling occurs (i.e. $\gamma = 1$). For $n=50$, the $NN$ test has a slightly lower power than the residual test. This difference diminishes as the sample size increases. Conversely, the $NN$ test outperforms when both the spatial range is very small ($\rho_Z = 0.02$) and when the magnitude of PS is very high ($\gamma = 2$). Fig.\ \ref{fig:power_hifreq_changen_rep} in the supplementary material demonstrates this. Under these conditions, very small clusters form. However, a different choice of bandwidth-selection method may improve the power of the residual test.  

Strong covariate effects in the sampling process hurts the power of all the tests. This can be seen in Fig \ref{fig:Power_n50_covar_rep} in the supplementary material. Interestingly, the $NN$ test is shown to be competitive across all settings tested, except one. When the spatial ranges of both the covariate $\textbf{w}(\textbf{s})$ and the field $Z(\textbf{s})$ are large and similar, the power of the $NN$ test is very low. Here, the residual test, with residuals computed from the fitted IPP, performs much better. The power is almost doubled that of the $NN$ test. In these settings, despite the $\textbf{w}(\textbf{s})$ and $Z(\textbf{s})$ arising from independent distributions, the empirical correlations between them in any given realisation may be large. Consequently, the $NN$ test may be unable to distinguish between the clustering due to $Z(\textbf{s})$, and the clustering due to the measured covariate $\textbf{w}(\textbf{s})$. This is because the $NN$ test, unlike the residual test, does not directly adjust for $\textbf{w}(\textbf{s})$. However, the IPP residual test is not always superior. When the $\rho_w$ is very low, the $NN$ test has higher power when $\rho_Z = 0.2$.

The performance of the tests are also assessed in settings where the response is non-Gaussian, and when the true sampling process $\cal{P}$ is not an IPP. (14) is replaced with a Poisson distribution and the true sampling process is set equal to a Hardcore process. Different radii of interactions are compared. The Hardcore process is purposefully chosen. Here, the use of nearest-neighbour distances to capture additional clustering is poor. Since the nearest neighbour distances are lower bounded, the contrast in their observed values will decrease as the radius of interaction increases. Conversely, estimates of the smoothed residuals will not be directly affected. Figure \ref{fig:HC_power} in the supplementary material clearly shows that the test based on the smoothed residuals far outperforms the $NN$ test when the radius of interaction is high. This demonstrates a clear need for the researcher to choose a measure of clustering that is suitable for the true sampling process. The power remains high for Poisson $f$. 

\section{Case Studies}

The ability of the test to detect PS in two real case studies is now demonstrated. These two datasets are chosen since the presence of PS within them has previously been shown in published work. In the first example, it is shown how researchers can easily detect positive PS and then search for a sufficient set of informative covariates $\textbf{x}(\textbf{s}, t)$ using the test. In the latter example, negative PS is detected. 

\subsection{Great Britain's black smoke monitoring network}

Annual concentrations of black smoke were obtained from the UK National Air Quality Information Archive (airquality.co.uk). Site locations and annual concentrations average concentrations of black smoke ($\mu g m^{-3}$) were obtained from the monitoring sites. Previous analyses demonstrated that the network had been preferentially sampled, with $h$ a positive monotonic function across the years of 1966-1996 \citep{watson2018general, shaddick2014case}. 

\begin{figure}[ht!]
    \centering
    \includegraphics[scale=0.25]{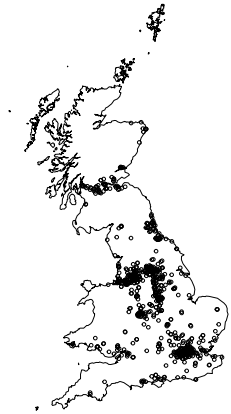}
    \caption{A plot of the locations of the black smoke monitoring sites in 1966.}
    \label{fig:blacksmoke1966}
\end{figure}

The analysis is restricted to the 1966 concentrations. For reasons outlined in \citep{shaddick2014case}, only sites that gave readings for at least 273 days of the year (75\% data capture) are considered. Fig.\ \ref{fig:blacksmoke1966} shows that it is apparent that sites were located in the industrial cities around London, the Midlands and the North West of England, with almost no sites present in the industry-free Scottish Highlands. Thus the point pattern displays clear evidence of clustering around regions expected to have high concentrations black smoke.

The R-INLA package with the SPDE approach is used to fit a standard geostatistical model \citep{rue2009approximate, lindgren2011explicit, lindgren2015bayesian}. Other methodologies, Bayesian or frequentist, could be used. As in \citet{watson2018general} the log-ratios of the concentrations are the choice of response. PC priors are placed on the approximate Matern field \citep{fuglstad2018constructing}. A prior probability that the spatial range is below 5km is set to 0.1, and a prior probability that the standard deviation of the field is above 3 is set as 0.1. The (mean-centered) posterior means of the log-transformed black smoke levels across $\cal S$ were then used as the `naive' $\hat{Z}(\textbf{s})$ values.

Gridded residential human population
count data with a spatial resolution of 1 km x 1 km were obtained for Great Britain. This was based on 2011 Census
data and 2015 Land Cover Map data from the Natural Environment Research Council Centre for
Ecology \& Hydrology \citep{UKgriddedpopdata}. As in \citet{watson2018general}, it is assumed that the relative population density across Great Britain remained approximately stable from 1966-2011. This is used as both a covariate $w(\textbf{s})$ for the null IPP sampling process and as a covariate $x(\textbf{s})$ for the observation process. Initially, the presence of PS is tested for without the population density covariate included in either process. This test is referred to as V1. Next, population density is controlled for as an informative covariate in both the observation and the null IPP sampling processes (it is found to be strongly associated with both). It is then investigated if PS remains. This test is referred to as V2.  


Estimates of $\hat{Z}(\textbf{s})$ in the V2 test are thus corrected for population density. Population density is included as a Bayesian spline to capture any nonlinear effects of population density on the observed black smoke $Y(\textbf{s})$. Mechanistically, including population density as an informative covariate in a model for black smoke concentration is reasonable. Localised sources of black smoke include the combustion of carbon-based fuels, with expected levels of combustion expected to increase with population density. If PS is no longer detected after this adjustment, then the population density has explained away the PS. Conversely, if PS is still detected, then standard `naive' methods may be biased even after controlling for population density. 

\begin{longtable}[c]{| c | c  c  c  c  c  c  c  c |}
 \caption{\label{tab:GBBS} A table of empirical p-values for the UK black smoke dataset for both the assumed homogeneous and inhomogeneous Poisson point process models.}\\
 \hline
  \em K&\em 1&\em 2&\em 3&\em 4&\em 5&\em 6&\em 7&\em 8\\
\hline
HPP P value V1 & 0.28 & 0.01 & 0.01 & 0.01 & 0.00 & 0.00 & 0.01 & 0.01 \\   
IPP P value V2 & 0.20 & 0.01 & 0.01 & 0.01 & 0.00 & 0.00 & 0.00 & 0.01 
\\ \hline
 \end{longtable}

Table \ref{tab:GBBS} shows the empirical pointwise p-values of the tests with changing $K$, under both of the assumed sampling mechanisms. Results are shown for both the $V1$ and $V2$ tests. The empirical p-values are the proportion of rank correlations in the 1000 Monte Carlo samples, that were more negative than observed in the data. Thus this is a 1-sided test. For $K > 1$, strong evidence is found against the null in favour of a positive monotonic $h$ under the $V1$ HPP test. This remains under the $V2$ IPP sampling process. Note that the p-values have not been adjusted for multiple testing.  

The insights gained from these tests are as follows. A `naive' model fit to the black smoke data without adjusting for population density may be biased due to PS being present. Whilst population density may explain some of the observed PS seen in the data, residual PS still remains after controlling for population density in a model for black smoke (IPP V2 result). Thus a sufficient set of PS-removing covariates has not yet been identified. Either this iterative process of finding a sufficient set of covariates must be continued, or a joint model should be considered as in \citet{watson2018general}.


\subsection{Galicia lead concentrations}

The second real world dataset consists of the concentrations of lead in moss samples collected across Galicia, northern Spain, in 1997 \citep{fernandez2000extended}. The concentration is measured in micrograms per gram of dry moss. The 1997 locations were previously shown to have been preferentially sampled in the landmark preferential sampling paper by \citet{diggle2010geostatistical}. In fact, in this example, a significant negative linear $h$ effect was found. Thus $h$ was found to be negative monotonic.

\begin{figure}[ht!]
    \centering
    \includegraphics[scale=0.25]{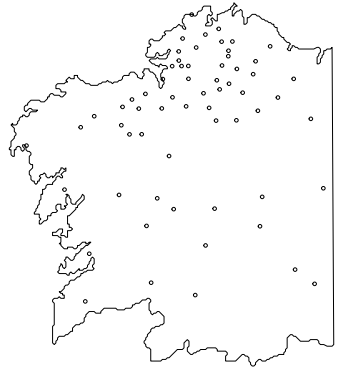}
    \caption{A plot of the 1997 sampled locations of lead concentrations in Galicia, northern Spain.}
    \label{fig:Galicia}
\end{figure}

Fig.\ \ref{fig:Galicia} shows a clear increase in the density of sampled locations in the northern half of Galicia. This half was found to have the lowest concentrations of lead. PC priors were again placed on the approximate Matern field. A prior probability that the spatial range is below 10km was set to 0.1, and the prior probability that the standard deviation of the field is above 3 was set to 0.1. The posterior mean log-transformed lead concentrations were used as the $\hat{Z}(\textbf{s})$ values. Empirical p-values were computed using 1000 Monte Carlo samples. The direction of the inequality was reversed this time, to test for negative PS.

\begin{longtable}[c]{| c | c  c  c  c  c  c  c  c |}
 \caption{\label{tab:Galicia} A table of empirical p-values for the Galicia dataset.}\\
 \hline
  \em K&\em 1&\em 2&\em 3&\em 4&\em 5&\em 6&\em 7&\em 8\\
\hline
P value & 0.01 & 0.03 & 0.06 & 0.06 & 0.07 & 0.08 & 0.08 & 0.09  \\   
\hline
\end{longtable}
\makeatletter
\global\advance\@colroom-90pt
\makeatother

The empirical pointwise p-values of this test are shown in Table \ref{tab:Galicia}. Even after considering Monte Carlo error, moderately strong evidence of negative PS exists, with the strongest evidence occurring at $K=1$. A researcher would now have to decide whether to pursue a sufficient set of covariates, or fit a joint model as in \citet{diggle2010geostatistical}.



\section{Concluding Remarks}

A fast and intuitive test has been presented for detecting preferential sampling (PS) in both geostatistical and discrete spatial data settings. The test is highly general; any preferred methodology for estimating a latent spatio-temporal process $Z(\textbf{s}, t)$ may be used. This includes both Bayesian and frequentist methods. In many situations, detected PS may be adequately described by a set of available informative covariates that are associated with both the sampling process and the observation process being measured. The test presented in this paper is able to identify such an informative covariate set. In cases where residual PS remains, even after controlling for a set of informative covariates, researchers can either seek a sufficient set of informative covariates, or seek a method that directly models the PS (e.g. a joint model). 

In this paper the properties and validity of the test were demonstrated through an extensive simulation study. The suitability of the test for real-world applications was then confirmed through the re-analysis of two previously published case studies. Both case studies had previously detected PS, and the test successfully replicated the findings of both. The power of the test to detect PS was shown to increase with: the spatial range (i.e. the inter-point distance required for observations to be approximately independent), the sample size, and the degree of PS. The power decreased dramatically with the inclusion of covariates, independent from the observation process, that had a strong effect on the sampling process.   

The biasing effects of PS on spatial prediction has been clearly demonstrated across a wide range of fields and can be severe in magnitude \citep{diggle2010geostatistical, watson2018general,pennino2019accounting, dinsdale2019modelling}. Thus, PS should not be ignored in spatial analyses. The test proposed in this paper has been shown to be suitable for assessing the presence of PS in most spatio-temporal analyses. Therefore, in cases where the sampling protocol is either unknown or known to be preferential, reporting the results from a PS test alongside any publication of spatio-temporal analyses should become standard practice. A user-friendly R package \textit{PStestR} is available for implementing the algorithm. No extra work or computation time is required. The package works seamlessly with many of the commonly used data types (e.g. from the \textit{sp}, \textit{sf} and \textit{spatstat} libraries \citep{sp1, sp2, sf, baddeley2015spatial}) and can perform both pointwise and rank envelope tests, to alleviate the multiple testing problem.

Three avenues of research should be pursued in the future. First, how to best capture the localised clustering in specific sampling settings should be explored. For example, in this paper it was shown that the nearest neighbour distance may be a poor choice for measuring the degree of clustering under certain sampling processes. How to optimise the power of the test in specific settings should be pursued. Next, eliminating the need for generating Monte Carlo samples from the null point process may be possible \citep{acosta2018effective}. We found that `effective-sample size'-adjusted rank correlation tests showed very poor performance. Convergence rates of the test could be as low as 10\% for specific simulation settings. Thus, we omitted the results. Further investigation here is warranted. Finally, adjustments are required for using this test in spatio-temporal applications where sampling locations are retained from one time step to the next. Environmental monitoring networks are an example. Here, the chosen network locations from one time step to the next are not independent. Additional work should be pursued to generalise the methods in this paper to such settings.   

\section*{Acknowledgments}
The author would like to extend his gratitude to Professor James V. Zidek FRSC, O.C. for his lively discussion and insightful feedback throughout.

\bibliographystyle{plainnat}
{\footnotesize
\bibliography{MonteCarlobib.bib}}

\newpage

\section{Supplementary material}

\subsection{More details on the simulation study}
For computational speed-ups, the R-INLA package is used for both the simulation and estimation of $Z(\textbf{s})$ \citep{rue2009approximate, lindgren2011explicit, lindgren2015bayesian}. A high resolution triangulation mesh (triangle lengths of 0.01) is defined for the SPDE approximation over the unit square $\cal S$, and linear interpolation is used to impute the values at any point $\textbf{s}$ within $\cal S$. For the priors on the Gaussian process, pc priors \citep{fuglstad2018constructing} are specified with prior probability of 0.1 that the spatial range is less than 0.1 and prior probability of 0.1 that the standard deviation is greater than 3. Gaussian errors on the responses $Y(\textbf{s}, t)$ are added, with a weakly informative Gamma(1, 5e$^-5$) distribution placed on the precision of the error distribution. This is done to reduce the risk of computational singularities. The $NN$ test is performed at the 5\% significance level using 19 Monte Carlo samples (i.e. $M = 19$). Each experimental setting is repeated 200 times.

Along with the $NN$ test outlined in algorithm 1, a Monte Carlo test using the rank correlations between estimates of $Z(\textbf{s})$ and estimates of the raw residuals of the assumed IPP under the null hypothesis is also compared. This may provide a more suitable test, when the assumed sampling mechanism is indeed a LGCP. Furthermore, such a test does not require a choice of $K$. To compute these residual values, an edge-corrected Gaussian kernel is used to smooth the raw residuals. The bandwidth is selected using leave-one-out cross-validation. This is performed using the \textit{spatstat} package \citep{baddeley2015spatial}. To compute the test, the $NN_k$ values are simply replaced with the smoothed residual values, evaluated at the point locations. We refer to this test as the residual test hereafter. It is interesting to assess the relative performance of the $NN$ test, given its generality across all point processes and to the discrete spatial setting.  


First, the Type 1 error of the PS tests are assessed in the simplest setting without any covariate effects (i.e. $\alpha_1 = 0$) or PS effects (i.e. $\gamma = 0$). Results from four tests are compared. The first two are residual tests. The first computes the p-value directly using a standard permutation approach under the (false) assumption that the pairs of residuals and estimates $\hat{Z}(\textbf{s})$ are an IID sample from some bivariate distribution. The positive spatial correlations due to the process $Z(\textbf{s})$ violate this assumption, with the magnitude of violation increasing with the spatial range $\rho_Z$. The second attempts to correct for this spatial correlation. By forming realisations from the estimated sampling process, the spatial correlation in $Z(\textbf{s})$ is accounted for. The third and fourth are $NN$ tests. Once again, comparisons are made between the permutation-based and the Monte Carlo-based approaches.

\begin{figure}[ht!]
    \centering
    \includegraphics[scale=0.6]{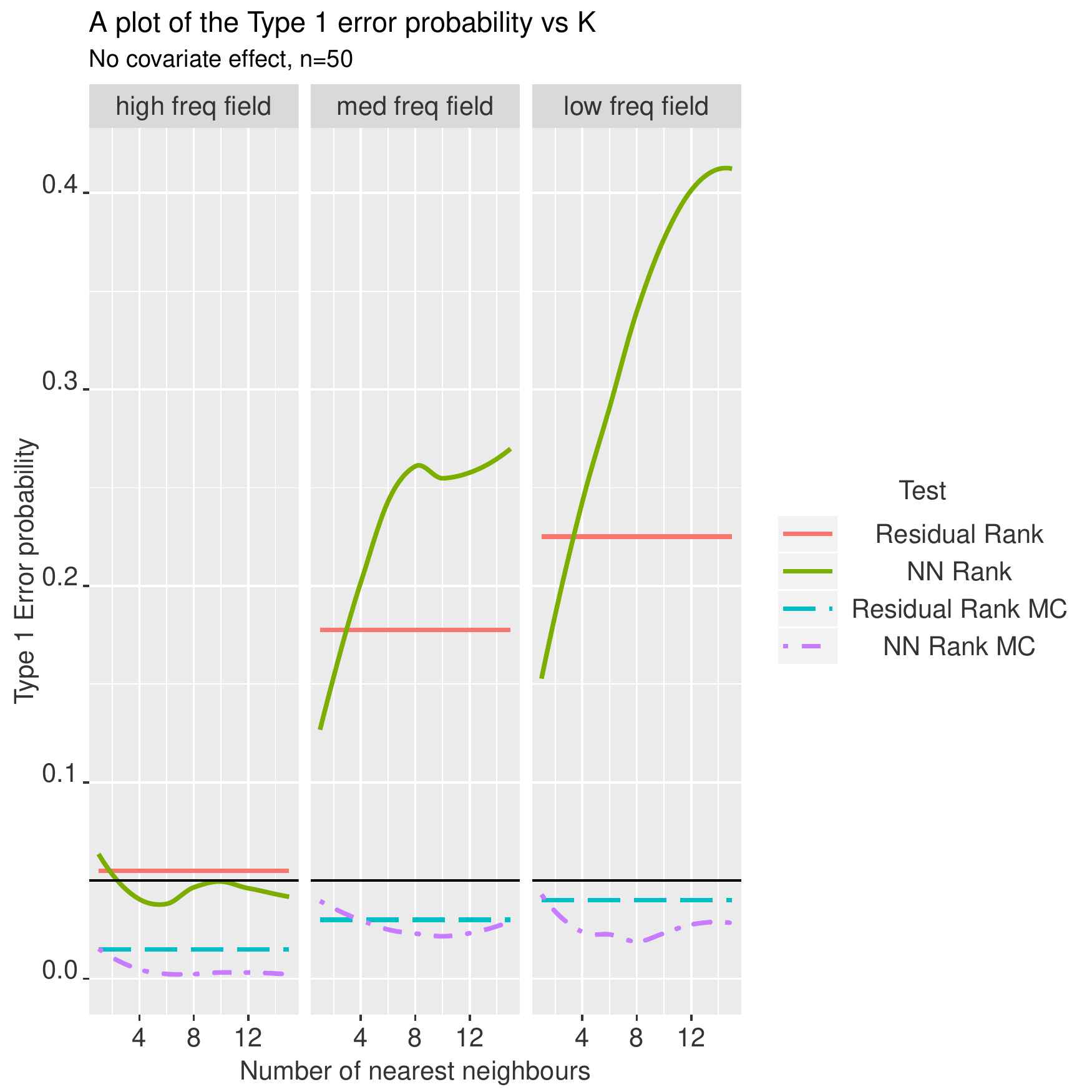}
    \caption{A plot of the Type 1 error for four tests. The three boxes show the results for $\rho_Z \in \{0.02, 0.2, 1\}$, from left to right respectively for a sample size of 50. The two `Residual' tests are computed using the kernel-density smoothed values of the residuals from the fitted homogeneous Poisson processes. Leave-one-out cross validation was used to select the bandwidth. The `NN' tests are those based on the $K$ nearest neighbour values. The suffix `MC' denotes the test has been computed from Monte Carlo realisations of the fitted point process.}
    \label{fig:Type1_n50_rep}
\end{figure}

\begin{figure}[ht!]
    \centering
    \includegraphics[scale=0.6]{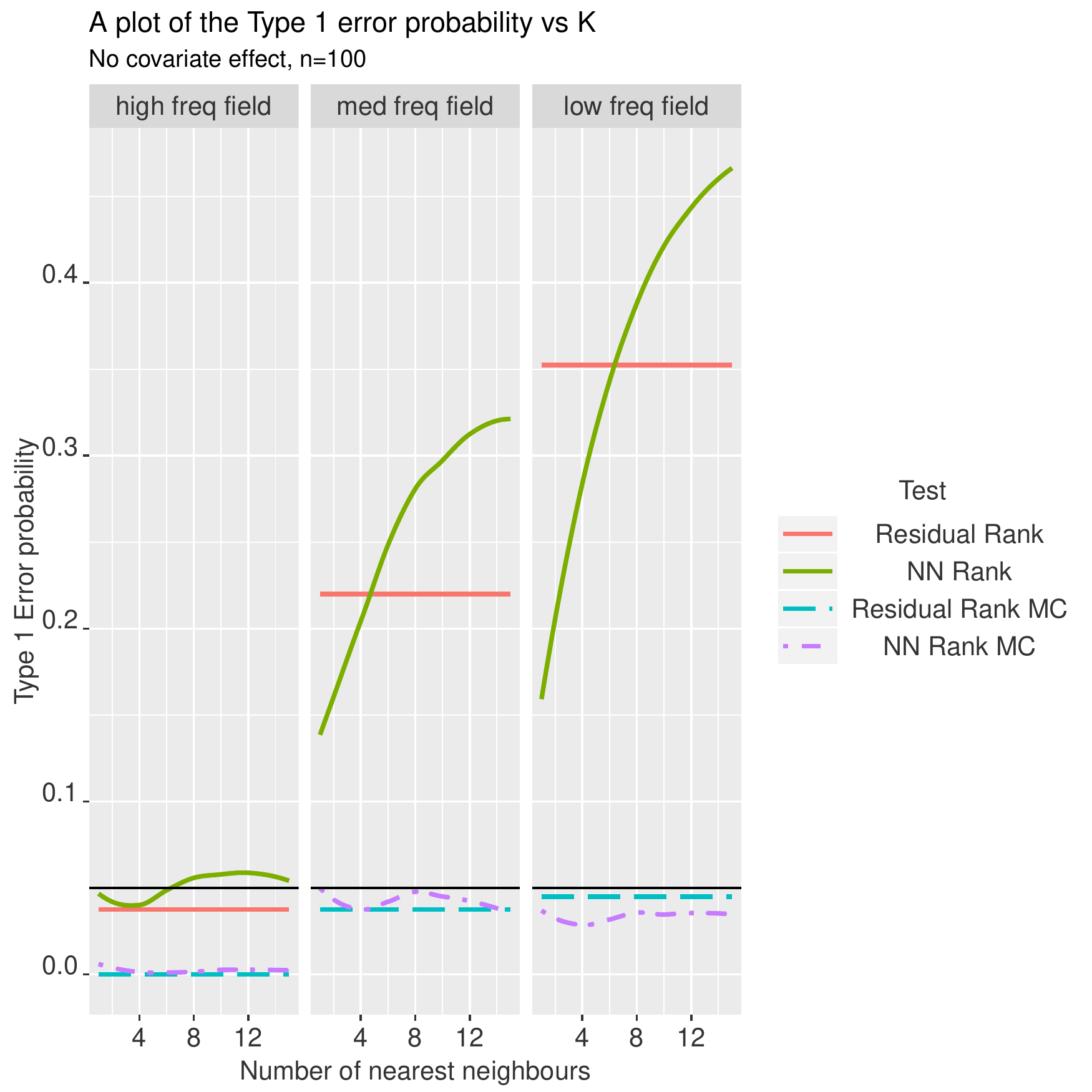}
    \caption{A plot of the Type 1 error for four tests. The three boxes show the results for $\rho_Z \in {0.02, 0.2, 1}$, from left to right respectively for a sample size of 100. The `Residual' tests are computed from the kernel-density smoothed values of the residuals from the fitted homogeneous Poisson processes. Leave-one-out cross validation was used to select the bandwidth. The `NN' tests are those based on the $K$ nearest neighbour values. The suffix `MC' denotes the test has been computed from Monte Carlo realisations of the fitted point process.}
    \label{fig:Type1_n100}
\end{figure}

Fig.\ \ref{fig:Type1_n50_rep} shows the results for $n=50$ across three increasing values of the spatial range $\rho_Z$ and across different numbers of nearest neighbours $K$. It is apparent that both Monte Carlo tests attain Type 1 error at or below the 5\% level. The two standard permutation tests attain a Type 1 error above the 5\% level, and this increases dramatically with $\rho_Z$. At the highest value of $\rho_Z = 1$, equal to the length of the domain $\cal S$, the Type 1 error can be higher than 40\%. The results for the very low value of $\rho_Z = 0.02$, demonstrate that the type 1 error approaches the nominal 5\% level when the spatial correlation approaches zero. This is due to the IID assumption becoming more reasonable as the $Z(\textbf{s})$ tends towards Gaussian white noise. When $\rho_Z = 0.02$, the prior distribution on the range parameter would reflect the case where a researcher incorrectly assumed spatially smooth data prior to the model-fitting. Fig.\ \ref{fig:Type1_n100} shows the results for $n=100$. It is apparent that the Type 1 error increases with sample size for the permutation tests, while the Monte Carlo tests remain bounded above by 0.05.

\begin{figure}[ht!]
    \centering
    \includegraphics[scale=0.6]{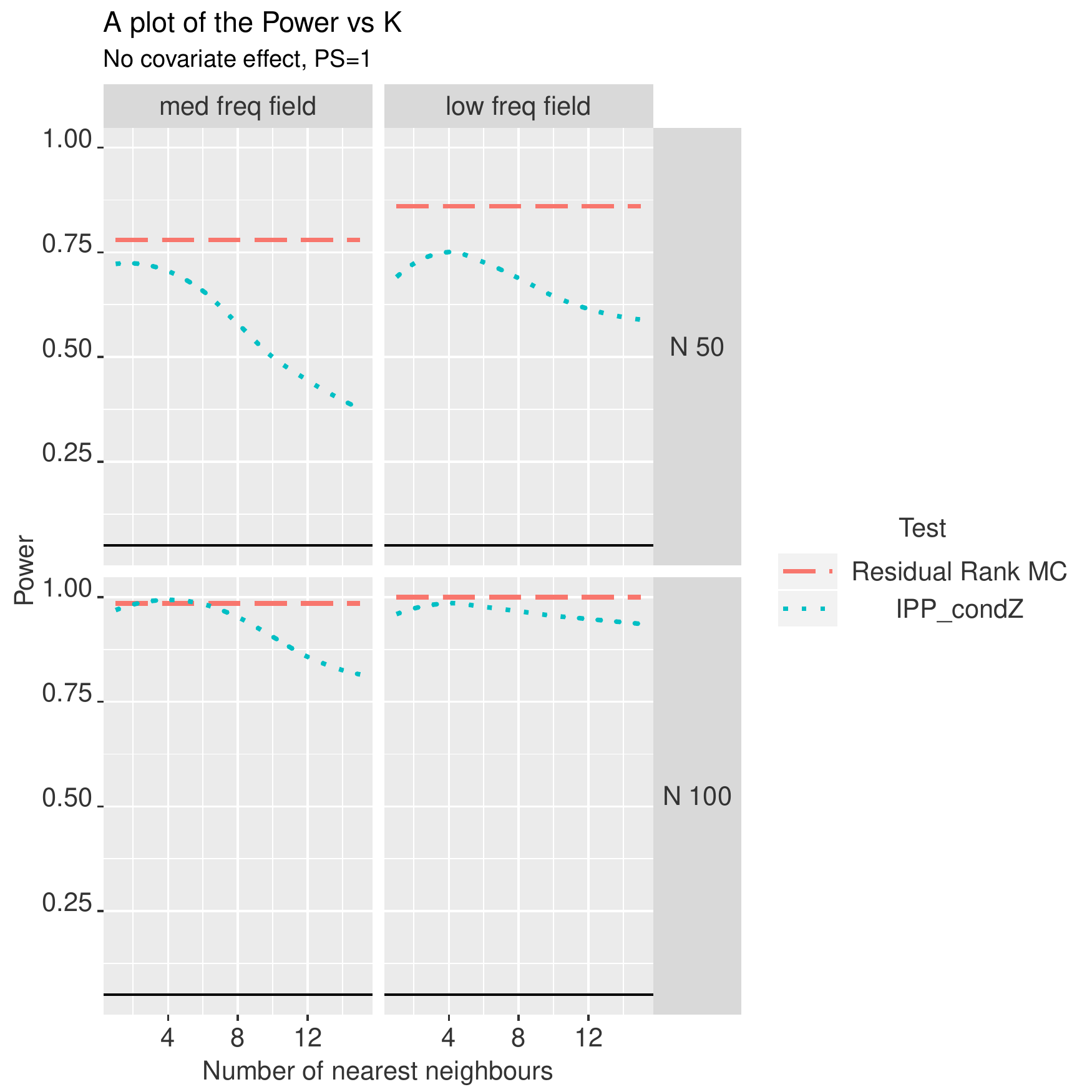}
    \caption{A plot of the Power for two tests when the PS parameter $\gamma$ equals 1. The two columns show the results for $\rho_Z \in {0.2, 1}$, from left to right respectively. The two rows show results for a sample size of 50 and 100 respectively. The `Residual' tests are computed using the kernel-density smoothed values of the residuals from the fitted homogeneous Poisson processes. Leave-one-out cross validation was used to select the bandwidth. The `NN' test is based on the $K$ nearest neighbour values. The suffix `MC' denotes the test has been computed from Monte Carlo realisations of the fitted point process.}
    \label{fig:Power_n50_rep}
\end{figure}

Next, the power of the two Monte Carlo tests to detect a PS effect when the alternative hypothesis is true is assessed. The behaviours of the tests are first investigated in the setting where no covariate effects exist (i.e. $\alpha_1 = 0$), but where moderate positive preferential sampling occurrs (i.e. $\gamma = 1$). All tests are performed with the two-sided alternative hypothesis, namely that $h$ is a monotonic function of $Z$ in either direction.

\begin{figure}[ht!]
    \centering
    \includegraphics[scale=0.6]{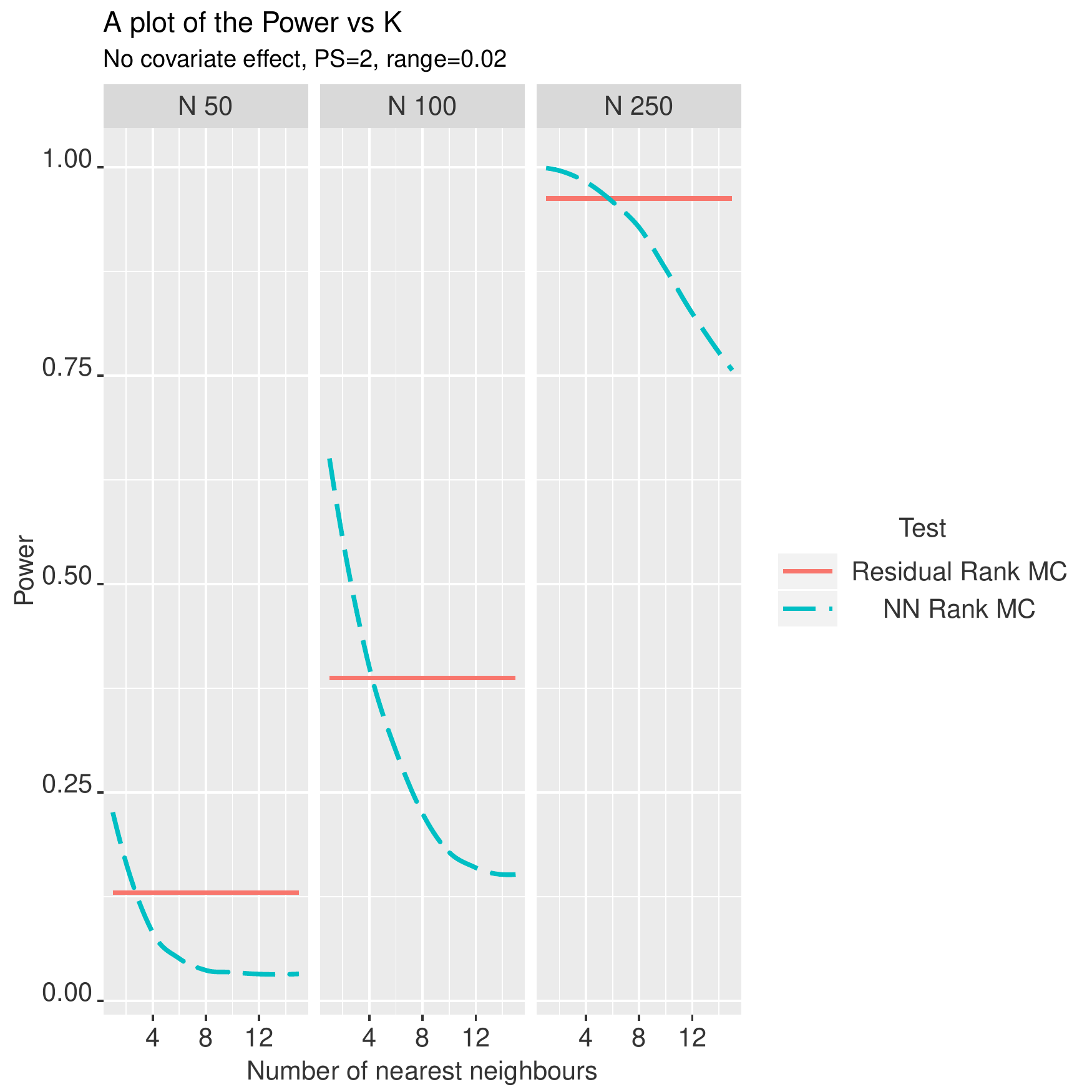}
    \caption{A plot of the Power for two tests when the PS parameter $\gamma$ equals 2 and $\rho_Z = 0.02$. The three columns show the results for the sample sizes 50, 100 and 250 from left to right respectively. The `Residual' tests are computed from the kernel-density smoothed values of the residuals from the fitted homogeneous Poisson processes. Leave-one-out cross validation was used to select the bandwidth. The `NN' test is based on the $K$ nearest neighbour values. The suffix `MC' denotes the test has been computed from Monte Carlo realisations of the fitted point process.}
    \label{fig:power_hifreq_changen_rep}
\end{figure}

Fig.\ \ref{fig:Power_n50_rep} shows the results for $n=50$, this time with spatial ranges of $\rho_Z \in \{0.2, 1\}$, again across $K \in \{1,...,15\}$. The power results for $\rho_Z = 0.02$ are omitted in Figure \ref{fig:power_hifreq_changen_rep}, since the power is consistently small (<0.1) for both. This demonstrates the need for the $Z(\textbf{s}, t)$ term to be spatially-smooth for the test to detect PS. It is clear that the power of the $NN$ test is sensitive to the choice of $K$ value, especially for smaller sample sizes. Interestingly the optimum power achieved by the $NN$ test with respect to $K$ depends upon both the spatial range of $Z$ and the sample size. The higher the spatial range of $Z$, and hence the smoother it is, the greater the value of $K$ that is required to optimise the power. The optimum choice of $K$ also increases with the sample size, since the number of realised points per cluster increases. For example, when $\rho_Z = 0.2$ and the sample size is 50, the test is optimised when $K=1$. This increases to $K=5$ when $\rho_Z = 1$ and the sample size is 100. Finally, for $n=50$ it appears that the $NN$ tests have a slightly lower power than the residual measure-based test.  

\begin{figure}[ht!]
    \centering
    \includegraphics[scale=0.6]{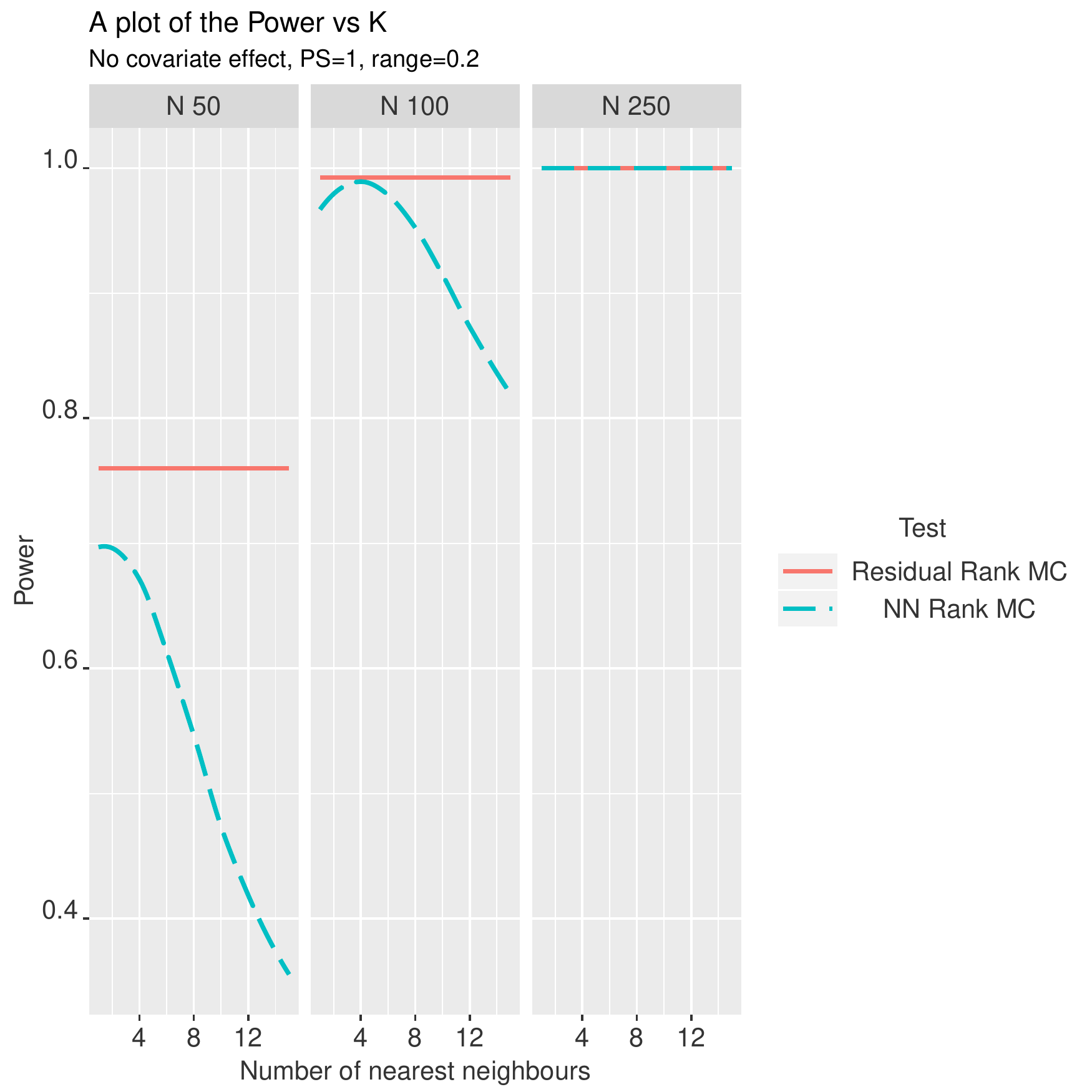}
    \caption{A plot of the Power for two tests when the PS parameter $\gamma$ equals 1 and $\rho_Z = 0.2$. The three columns show the results for the sample sizes 50, 100 and 250 from left to right respectively. The two rows show results for a sample size of 50 and 100 respectively. The `Residual' test denotes the kernel-density smoothed values of the raw residuals from the homogeneous Poisson process. Leave-one-out cross validation was used to select the bandwidth. The `NN' tests are those based on the $K$ nearest neighbour values. The suffix `MC' denotes the test has been computed from Monte Carlo realisations of the fitted point process.}
    \label{fig:power_midfreq_changen}
\end{figure}

Next, the spatial range is fixed to be very small ($\rho_Z$ = 0.02), and the magnitude of PS is fixed to be very high ($\gamma = 2$). This set-up leads to very small clusters to form when $\gamma \neq 0$. The joint effects of sample size and $K$ on the power of the $NN$ test to detect PS is then demonstrated. Additionally, the power of the $NN$ test is compared to the residual test. Three plots are shown to present the power vs. $K$ in Fig.\ \ref{fig:power_hifreq_changen_rep}. From left to right, these show sample sizes of $50, 100$ and $250$. For small sample sizes ($n \in \{50, 100\}$), both tests have low power to detect PS as expected. Interestingly however, the $NN$ test outperforms the smoothed residual test for all three sample sizes, achieving maximum powers of 0.21, 0.65 and 1 at $K=1$ compared with 0.14, 0.40 and 0.97 for the residual test. Furthermore, the power of the $NN$ test attains it maximum at $K=1$, before dramatically diminishing to 0 as $K$ increases. Fig.\ \ref{fig:power_midfreq_changen} shows the equivalent plots for $\rho_Z = 0.2$. Here, the $NN$ test is no longer more powerful, with the residual test performing better at $n = 50$. Note that the performance of the residual test may improve with a different choice of bandwidth-selection method.  


Results are now presented for the case when a unique covariate effect exists for the sampling process. The magnitudes of the covariate effect and the PS effect are both set to 1 (thus $\alpha_1 = \gamma = 1$). The spatial range of the covariate effect is varied ($\rho_w \in \{1, 0.02\}$). The results on the power of three different tests to detect PS are shown. As before, the first two are the kernel-smoothed residual and $NN$ rank tests. The third test is a rank test using kernel-smoothed estimated residuals, but this time using residuals computed from an incorrectly specified point process fit to the points. This is chosen to be a homogeneous Poisson process (HPP hereafter). Note that the Monte Carlo realised points $S_t^m$ still come from the null IPP, fitted to the original observations $S_t$. Thus the Monte Carlo sampled realisations still come from the correct data generating mechanism (correct up to parameter estimation error). Unlike the residuals from the first test, these residuals do not adjust for the covariate effect. The purpose of this comparison is to see if any improvements in the power of the test can be attained by considering computed quantities that directly adjust for any covariate effects. 

\begin{figure}[ht!]
    \centering
    \includegraphics[scale=0.6]{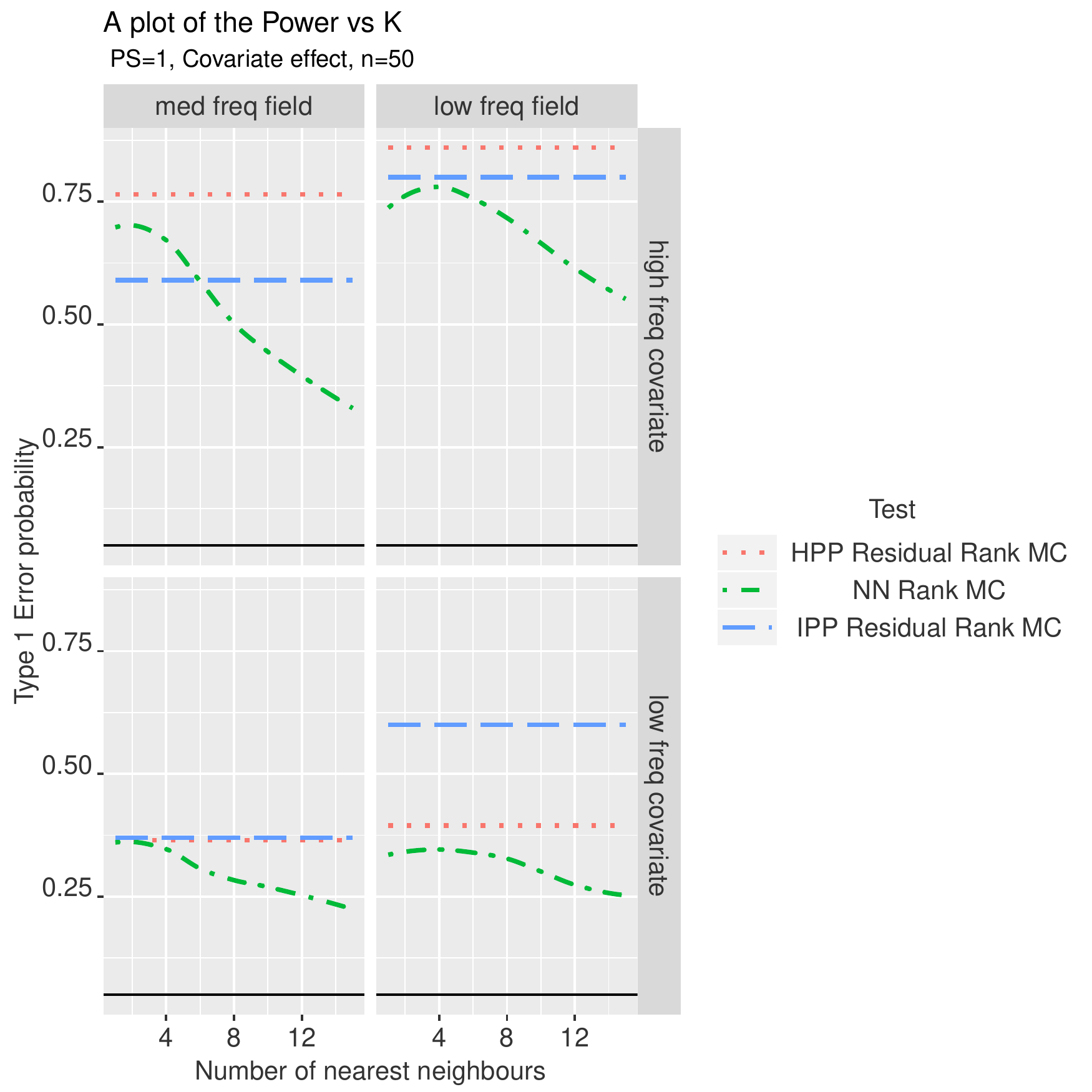}
    \caption{A plot of the Power for two tests when the PS parameter $\gamma$ equals 1, the covariate effect $\alpha_1$ equals 1 and when the sample size is 50. The two columns show the results for the spatial range $\rho_Z \in \{0.2, 1\}$ from left to right respectively. The two rows show the results for the spatial range of the covariate $\rho_w \in \{0.02, 1\}$ from top to bottom respectively. The two `Residual' tests are computed from the kernel-density smoothed values of the raw residuals from the fitted Homogeneous (HPP) and Inhomogeneous Poisson processes (IPP). Leave-one-out cross validation was used to select the bandwidth. The `NN' test is based on the $K$ nearest neighbour values. The suffix `MC' denotes the test has been computed from Monte Carlo realisations of the fitted point process. }
    \label{fig:Power_n50_covar_rep}
\end{figure}

The spatial range of the covariate field is changed for the following reason. When the spatial ranges of both the covariate field $\textbf{w}(\textbf{s})$ and the underlying spatial field $Z(\textbf{s})$ are large and similar, then the magnitude of the empirical correlation of a single realisation of the two fields may be high. This is despite their realisations arising from independent distributions \citep{hanks2015restricted}. A possible consequence of this is that the tests may be unable to distinguish between clustering due to an unknown process $Z$, and clustering due to the measured covariate $\textbf{w}(\textbf{s})$. This may affect the ability of tests to detect preferential sampling, when their computed quantities are not adjusted for the effects of covariates. The rank test of the residuals from the correctly specified IPP model is the only test that directly adjusts for the covariate effects. 

Fig \ref{fig:Power_n50_covar_rep} presents a complex interaction of different factors. When the covariate field has very low spatial range (i.e. $\rho_w = 0.02$), and hence has high frequency, negligible correlation can exist between $Z(\textbf{s})$ and $w(\textbf{s})$. Consequently, no gains in power are seen when tests use covariate-adjusted measures of clustering relative to when they use unadjusted measures. However, when both the covariate $w$ and underlying field $Z$ are very smooth (i.e. $\rho_w = 1, \rho_Z = 1$), large increases in power are seen with the covariate-adjusted measure of clustering. The power increases to 0.57 compared with 0.40 and 0.33 for the HPP residual and the $NN$ methods respectively. The results are similar for $n = 250$ (see Fig.\ \ref{fig:Power_n250_covar} in the supplementary material). In conclusion then, in cases where the spatial ranges of informative covariates are large and similar in size to the underlying $Z(\textbf{s})$, computed quantities other than $NN_k$ should be considered to improve the power to detect PS.


Finally, the performance of the tests are assessed in settings where the response is non-Gaussian, and when the true sampling process is not an IPP. The $Y(\textbf{s})$ values now take the form of counts and (13) is replaced with a Poisson distribution. The log-transformed mean at location $\textbf{s} \in \cal S$ is set equal to the random field $Z(\textbf{s})$ plus a constant intercept of 2. The intercept is chosen to ensure non-zero counts occur often. The true sampling process is set equal to a Hardcore process with two different radii of interactions, denoted $R$, compared (0.025 and 0.05). Under these two processes, points within $S_t$ cannot be sampled closer than a distance apart of 0.025 or 0.05 respectively. With $n=100$, these two constraints enforce moderate and strict levels of regularization of the points respectively, violating the IPP assumption of no inter-point interaction. 

The Hardcore process is chosen to highlight the fact that the choice of nearest-neighbour distances to capture additional clustering will be poor in some settings. Here, due to the nearest neighbour distances being lower bounded, the contrast in their observed values will decrease as $R$ is increased. Estimates of the smoothed residuals are not directly affected by an increase in $R$ and hence the residual test is expected to far outperform the NN test as $R$ increases.  

After sampling the data, the tests under two scenarios are compared. The first considers the case where the researcher assumes the correct Hardcore process sampling mechanism for the Monte Carlo realisations of $S_t$. The second considers the case when the researcher misspecifies it as an IPP (i.e. assumes no inter-point interaction). 

\begin{figure}[ht!]
    \centering
    \includegraphics[scale=0.6]{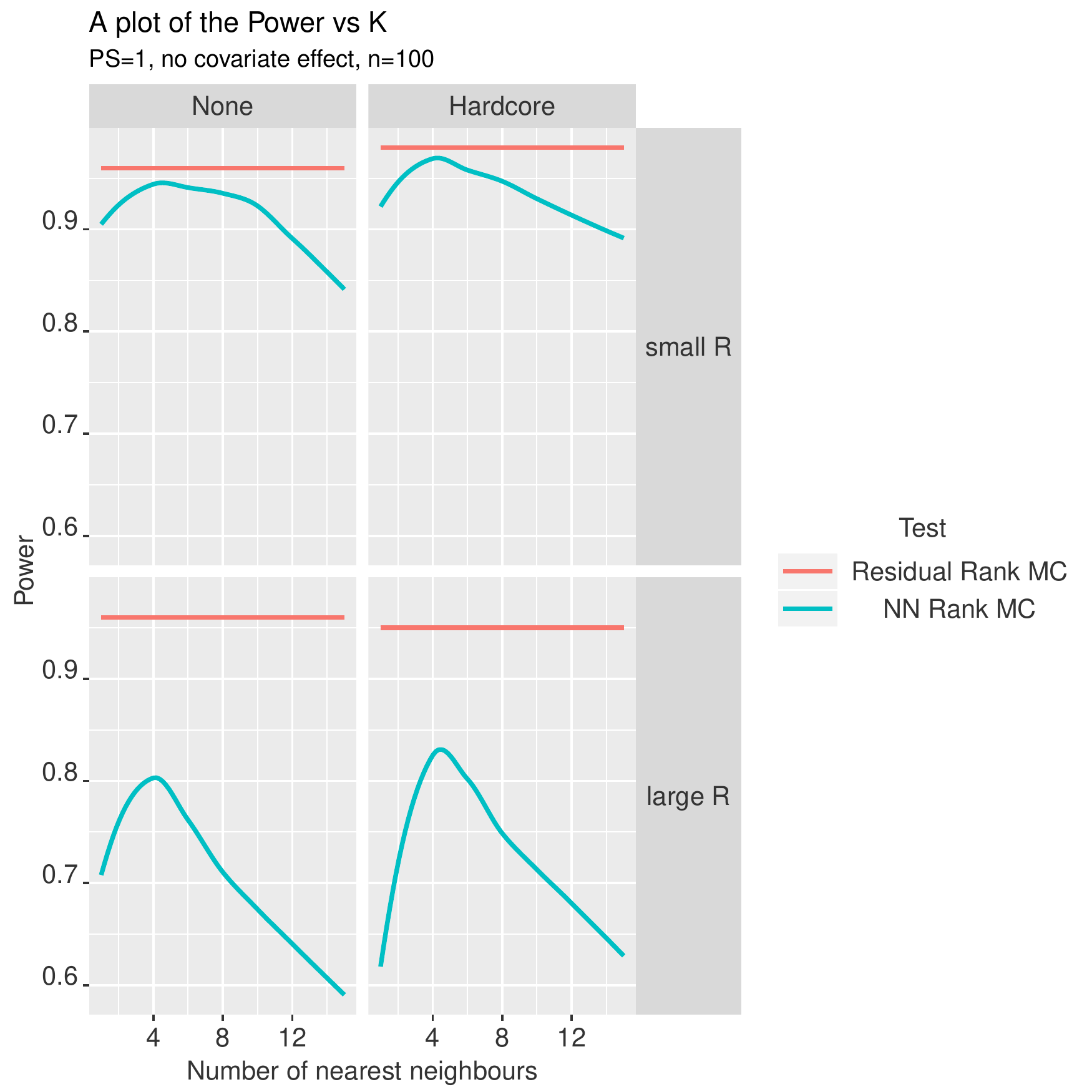}
    \caption{A plot of the Power for two tests when the true sampling process is a Hard Core point process, with PS parameter $\gamma$ equals 1 and $\rho_Z = 1$. The two columns show the results when a Poisson process model (`None'), and when a Hard Core process are fitted and then used for Monte Carlo sampling. From top to bottom, the rows denote the case where the true radius of interaction for the Hard Core process equals 0.025 and 0.05. The `Residual' test is computed using kernel-density smoothed values of the raw residuals from the fitted point process. Leave-one-out cross validation was used to select the bandwidth. The `NN' tests are those based on the $K$ nearest neighbour values. The suffix `MC' denotes the test has been computed from Monte Carlo realisations of the fitted point processes.}
    \label{fig:HC_power}
\end{figure}

The results from this simulation study, repeated 100 times, are shown in Fig.\ \ref{fig:HC_power}. As expected, the residual tests far outperform the NN test when the radius of interaction is 0.05. This is due to the lack of contrast in the nearest neighbour distances that leads to a reduction in the power of the NN test. When the radius of interaction is 0.025, the contrast in the nearest neighbour distances is restored and both methods perform very well again. In this case, the power exceeds 0.95 when the correct Hard Core process is fitted. Interestingly, the residual test that use the raw residuals from the correctly specified Hard Core processes perform no better than the residual test that use raw residuals of the incorrectly specified HPP. On the other hand, the performance of the NN test improves when the class of point process is correctly specified.

\begin{figure}[ht!]
    \centering
    \includegraphics[scale=0.6]{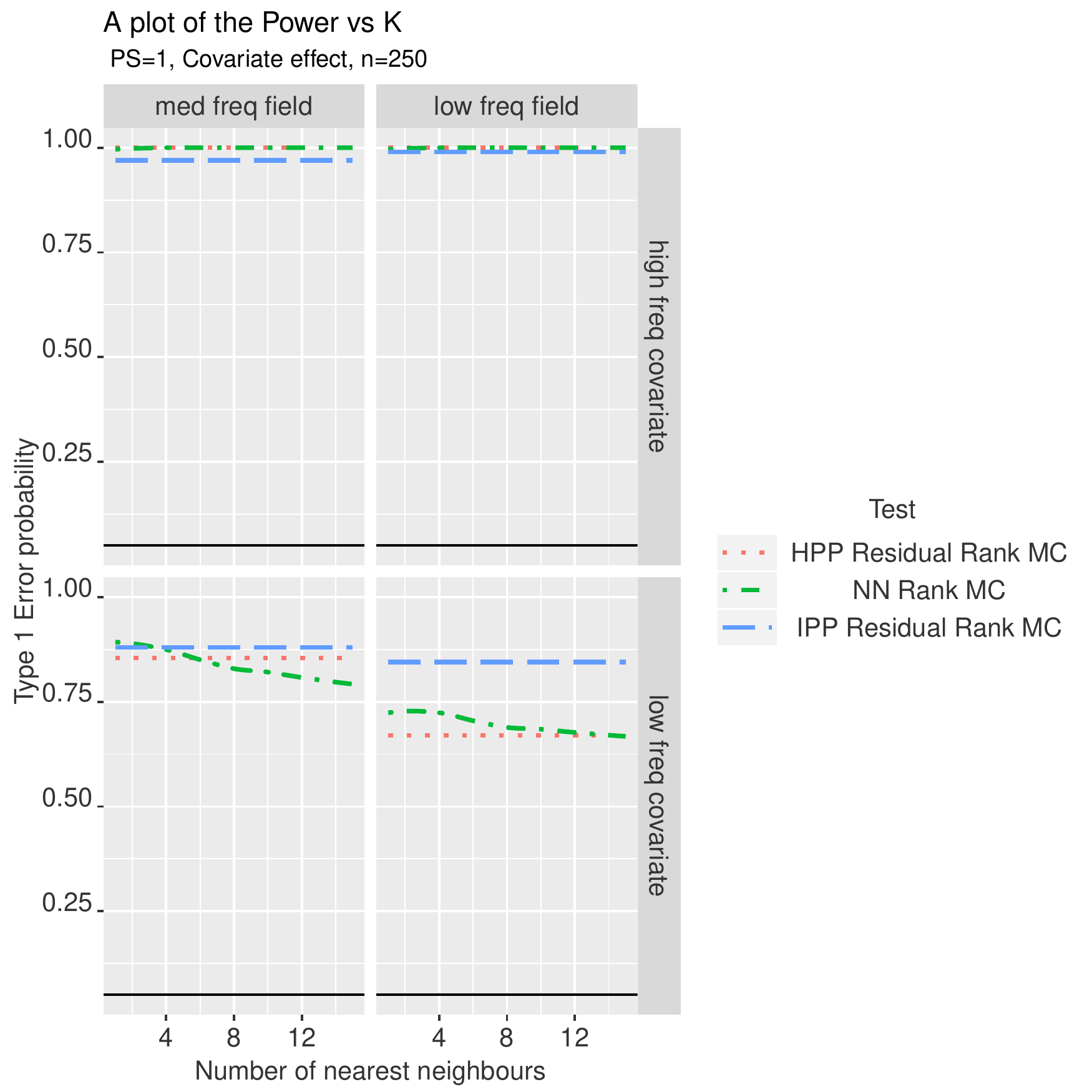}
    \caption{A plot of the Power for two tests when the PS parameter $\gamma$ equals 1, the covariate effect $\alpha_1$ equals 1 and when the sample size is 250. The two columns show the results for the spatial range $\rho_Z \in \{0.2,1\}$ from left to right respectively. The two rows show the results for the spatial range of the covariate $\rho_w \in \{0.02, 1\}$ from top to bottom respectively. The `Residual' test is computed using kernel-density smoothed values of the raw residuals from the fitted Homogeneous and Inhomogeneous Poisson processes. Leave-one-out cross validation was used to select the bandwidth. The `NN' test is based on the $K$ nearest neighbour values. The suffix `MC' denotes the test has been computed from Monte Carlo realisations of the fitted point process. }
    \label{fig:Power_n250_covar}
\end{figure}

\end{document}